\documentclass[prd,twocolumn,nofootinbib,showpacs,10pt,superscriptaddress]{revtex4-1}
\usepackage{textcomp}
\usepackage{amsmath}
\usepackage{amsfonts}
\usepackage[bbgreekl]{mathbbol}
\usepackage{calc}
\usepackage{mathrsfs}
\usepackage{amssymb}
\usepackage{amsmath}
\usepackage{array}
\usepackage{color}
\usepackage{bbm}
\usepackage{graphicx}
\usepackage{hyperref}
\usepackage{color}
\usepackage[usenames,dvipsnames,svgnames,table]{xcolor}

\renewcommand{\P}{\mathcal{P}}
\newcommand{\res}{\mathcal{R}}
\renewcommand{\S}{{\rm S}}

\newcommand{\lmax}{{ l_{\rm max}}}
\newcommand{\mmax}{{m'_{\rm max}}}

\newcommand{\beq}{\begin{equation}}
\newcommand{\eeq}{\end{equation}}
\renewcommand{\O}{\mathcal{O}}

\newcommand{\sub}[1]{{}_{\!\! #1}}

\begin{document}
\title{Second-order perturbation theory: the problem of infinite mode coupling} 
\author{Jeremy Miller}
\affiliation{Mathematical Sciences and STAG Research Centre, University of 
Southampton, Southampton, SO17 1BJ, United Kingdom}
\author{Barry Wardell}
\affiliation{School of Mathematical Sciences and Complex \& Adaptive Systems 
Laboratory,
University College Dublin, Belfield, Dublin 4, Ireland}
\affiliation{Department of Astronomy, Cornell University, Ithaca, NY 14853, USA}
\author{Adam Pound}
\affiliation{Mathematical Sciences and STAG Research Centre, University of 
Southampton, Southampton, SO17 1BJ, United Kingdom}
\date{\today}

\begin{abstract}
Second-order self-force computations, which will be essential in modeling 
extreme-mass-ratio inspirals, involve two major new difficulties that were not 
present at first order. One is the problem of large scales, discussed in [Phys. 
Rev. D 92, 104047 (2015)]. Here we discuss the second difficulty, which occurs 
instead on small scales: if we expand the field equations in spherical harmonics, then because the first-order field contains a 
singularity, we require an arbitrarily large number of first-order modes to 
accurately compute even a single second-order mode. This  is a generic feature 
of nonlinear field equations containing singularities, allowing us to study it 
in the simple context of a scalar toy model in flat space. Using that model, we 
illustrate the problem and demonstrate a robust strategy for overcoming it.
\end{abstract}

\maketitle

\section{Introduction and summary}

Gravitational self-force theory~\cite{Poisson-Pound-Vega:11,Pound:15a,Harte:15} 
has proven to be an important tool in efforts to model compact binary 
inspirals. It is currently the only viable method of accurately modeling 
extreme-mass-ratio inspirals (EMRIs)~\cite{Barack:09,Amaro-Seoane-etal:14}, it 
is a potentially powerful means of modeling intermediate-mass-ratio inspirals, 
and by interfacing with other methods, it can even be used to validate and improve models of 
comparable-mass binaries~\cite{Blanchet-etal:10b,Akcay-etal:12,Blanchet-etal:14,LeTiec:14, Bernuzzi-etal:14,Damour-etal:16}. However, the self-force model is 
based on an asymptotic expansion in the limit $m/M\to0$, where $m$ and $M$ are 
the two masses in the system. The model's accuracy is hence limited by the 
perturbative order at which it is truncated. Unfortunately, although numerous concrete 
self-force computations of binary dynamics have  been performed (see the reviews~\cite{Barack:09,Poisson-Pound-Vega:11,Wardell:15} and Refs.~\cite{Colleoni-etal:15,Bini-etal:15,Osburn-Warburton-Evans:16, Hopper-etal:16, Kavanagh-etal:16, vandeMeent:16,Akcay-Dempsey-Dolan:16} for some more recent examples), 
until now they have been restricted to first perturbative order, limiting their capacity 
to assist other models and rendering them insufficiently accurate to model 
EMRIs~\cite{Hinderer-Flanagan:08}.

In recent years, substantial effort has gone into overcoming this 
limitation~\cite{Rosenthal:06a,Rosenthal:06b,Detweiler:12,Pound:12a, Gralla:12, 
Pound:12b, Pound-Miller:14, Pound:15a, Pound:15b,Pound:15c,Pound:16}. The 
foundations of second-order self-force theory are now 
established~\cite{Pound:12a,Pound:12b,Gralla:12,Pound:15a}, the key analytical 
ingredients are in place~\cite{Pound-Miller:14}, and at least in some 
scenarios, practical formulations of the second-order field equations have been 
developed~\cite{Warburton-Wardell:14,Wardell-Warburton:15,Pound:16}. However, 
concrete solutions to the field equations have remained elusive.

There have been two major obstacles to finding these solutions. The first is 
the problem of large scales, described in Ref.~\cite{Pound:15c}, which 
manifests in spurious unbounded growth and ill-defined retarded integrals. As 
demonstrated in a simple toy model in Ref.~\cite{Pound:15c}, this obstacle can 
be overcome by utilizing multiscale and matched-expansion techniques; full 
descriptions of these techniques in the gravitational problem will be given in 
future papers. The second major obstacle arises in the opposite extreme: rather 
than a problem on large scales, it is a problem on small ones. 

To introduce the problem, we refer to the Einstein equations through second 
order, which we can write as
\begin{align}
\delta G_{\mu\nu}[h^1] &= 8\pi T_{\mu\nu},\label{EFE1-4D}\\
\delta G_{\mu\nu}[h^2] &= -\delta^2 G_{\mu\nu}[h^1,h^1].\label{EFE2-4D}
\end{align}
Here the metric has been expanded as 
$g_{\mu\nu}+(m/M)h^1_{\mu\nu}+(m/M)^2h^2_{\mu\nu}+\O(m^3)$; $T_{\mu\nu}$ is the 
stress-energy of a point particle, representing the leading approximation to 
the smaller object $m$ on the background $g_{\mu\nu}$; $\delta G_{\mu\nu}$ is 
the linearized Einstein tensor (in some appropriate gauge~\cite{Pound:15b}); and $\delta^2 G_{\mu\nu}[h^1,h^1]$ is the 
second-order Einstein tensor, which has the schematic form $h^1\partial^2 
h^1+\partial h^1\partial h^1$. Because $h^1_{\mu\nu}$ is singular at the 
particle, Eq.~\eqref{EFE2-4D} is only valid at points away from the particle's 
worline~\cite{Pound:12b}, but that suffices for our purposes here. 

Equations~\eqref{EFE1-4D}--\eqref{EFE2-4D} can in principle be solved in four 
dimensions (4D). However, in practice it is desirable to reduce their dimension 
by decomposing them into a basis of harmonics. For illustration let us use some 
 basis of tensor harmonics $Y^{ilm}_{\mu\nu}$; here we use the notation of 
Barack-Lousto-Sago~\cite{Barack-Lousto:05,Barack-Sago:07}, with 
$i=1,\ldots,10$, but the particular choice of basis, whether spherical or 
spheroidal (for example), is immaterial. We have 
\beq
h^n_{\mu\nu}=\sum_{ilm}h^n_{ilm}Y^{ilm}_{\mu\nu}\label{hilm}
\eeq
 and
\begin{align}
\delta G_{ilm}[h^1] &= 8\pi T_{ilm},\label{EFE1}\\
\delta G_{ilm}[h^2] &= -\delta^2 G_{ilm}[h^1,h^1].\label{EFE2}
\end{align}
Now consider the source term $\delta^2 G_{ilm}$. Substituting the 
expansion~\eqref{hilm} into $\delta^2 G_{\mu\nu}$ leads to a mode-coupling 
formula with the schematic form
\beq
\delta^2 G_{ilm} = 
\sum_{\substack{i_1l_1m_1\\i_2l_2m_2}}\mathscr{D}^{i_1l_1m_1i_2l_2m_2}_{ilm}[
h^1_{i_1l_1m_1},h^1_{i_2l_2m_2}],\label{ddGilm}
\eeq
where $\mathscr{D}^{i_1l_1m_1i_2l_2m_2}_{ilm}$ is a bilinear differential 
operator (given explicitly in Ref.~\cite{Pound:16}). A single mode $\delta^2 
G_{ilm}$ is an infinite sum over first-order modes $h^1_{ilm}$. If $h^1_{ilm}$ 
falls off sufficiently rapidly with $l$, then the summation poses no problem. 
However, if $h^1_{ilm}$ falls off slowly with $l$, then the summation is 
potentially intractable. This is precisely the situation near the 
point-particle singularity in Eq.~\eqref{EFE1}. $h^1_{\mu\nu}$ behaves 
approximately as a Coulomb field, blowing up as $\sim1/\rho$, where $\rho$ is a 
spatial distance from the particle. The individual modes 
$h^1_{ilm}Y^{ilm}_{\mu\nu}$, after summing over $m$, then go as $\sim 
l^0$ on the particle~\cite{Barack:09,Wardell-Warburton:15}, not decaying at 
all; at points \emph{near} the particle, the decay is arbitrarily slow.

This behavior can be understood from the textbook example of a Coulomb field 
$\varphi$ in flat space. For a static charged particle at radius $r_0$, the 
field's modes behave as $\varphi_{lm}Y_{lm}\sim(r_</r_>)^l$, where $r_<:={\rm 
min}(r_0,r)$ and $r_>:={\rm max}(r_0,r)$. On the particle, where $r=r_0$, we 
have $\varphi_{lm}Y_{lm}\sim l^0$. At any point $r\neq r_0$, we have 
exponential decay with $l$, but that decay is arbitrarily slow when $r\approx 
r_0$. Extrapolating this behavior to the gravitational case~\eqref{ddGilm}, we 
can infer that unless the coupling operator 
$\mathscr{D}^{i_1l_1m_1i_2l_2m_2}_{ilm}$ introduces rapid decay (which it does 
not), we are faced with the following tenuous position: \emph{to obtain a 
single mode of the second-order source near the particle, we must sum over an 
arbitrarily large number of first-order modes}.

In this paper, we explicate this problem and present a robust, broadly 
applicable method of surmounting it. Rather than  facing the challenge  head-on 
in gravity, we retreat to the same flat-space scalar toy model as was used in 
Ref.~\cite{Pound:15c}. In place of the gravitational field 
equations~\eqref{EFE1}--\eqref{EFE2}, we consider the field equations
\begin{align}
\Box\varphi_1 &= -4\pi\varrho,\label{phi1}\\
\Box\varphi_2 &= t^{\alpha\beta}\partial_{\alpha}\varphi_1\partial_\beta 
\varphi_1 =: S\label{phi2}.
\end{align}
Here, in Cartesian coordinates $(t,x^i)$, 
$\Box=-\partial_t^2+\partial^i\partial_i$ is the flat-space d'Alembertian, 
\beq\label{rho}
\varrho := \frac{\delta(x^i-x^i_p)}{dt/d\tau}
\eeq
is a point charge distribution moving on a worldline $x_p^\mu(t)=(t,x^i(t))$ 
with proper time $\tau$, and $t^{\mu\nu}:={\rm diag}(1,1,1,1)$. With our chosen 
source terms, the first-order field $\varphi_1$ mimics the behavior of 
$h^1_{\mu\nu}$, and the second-order source $S$ mimics the behavior of 
$\delta^2 G_{ilm}$. 

Like Eq.~\eqref{EFE2},  Eq.~\eqref{phi2} is well defined only at points off the 
worldline. To solve it globally, one would have to rewrite it as 
$\Box\varphi_2^\res=S-\Box\varphi^\P_2$~\cite{Pound:15c}, where $\varphi^\P_2$ 
is an analytically determined, singular ``puncture'' that guarantees the total 
field has the correct physical behavior near the particle, and 
$\varphi_2^\res:=\varphi_2-\varphi_2^\P$ is the regular ``residual'' difference 
between the total field and the puncture. However, here we only wish to address 
the preliminary question: \emph{given the spherical harmonic modes of 
$\varphi_1$, how can we accurately compute the modes of $S$}? Once that 
question is answered, the same method can be carried over directly to the 
gravitational case to compute the source $\delta^2G_{ilm}$, and 
Eq.~\eqref{EFE2} can then be solved via a puncture scheme of the sort described 
in Refs.~\cite{Warburton-Wardell:14,Wardell-Warburton:15} (see also 
Ref.~\cite{Pound:15c}).

Before describing the technical details of our computations, we summarize the 
problem, our strategy for overcoming it, and our successful application of that 
strategy. For simplicity, we fix the particle on a circular orbit of radius 
$r_0$. The modes  $\varphi^{\rm ret}_{lm}$ of the first-order retarded field 
are then easily found; they are given by Eqs.~\eqref{phi1lm!=0} and 
\eqref{phi1l0}. (To streamline the notation, we shall omit the subscript 1 on 
first-order fields.) From those modes, one can naively attempt to compute the 
modes $S_{lm}$ of the source using an analog of Eq.~\eqref{ddGilm}, given 
explicitly by Eq.~\eqref{Slm} below. Figure~\ref{Fig:slow convergence} shows 
the failure of this direct computation in the case of the monopole mode 
$S_{00}$. The higher the curve in the figure, the greater the number of 
first-order modes included in the sum, up to a maximum $l=\lmax$. Although the 
convergence is rapid at points far from the particle, it becomes arbitrarily 
slow near the particle's radial position $r_0$. In principle, this obstacle 
could be overcome with brute force, simply adding more modes until we achieve 
some desired accuracy at some desired nearest point to the particle. However, 
that relies on having all the modes of the retarded field at hand; in most 
applications of the self-force formalism, the retarded field modes are found 
numerically, and the number of modes is limited by practical computational 
demands. Hence, we should rephrase the question from the previous paragraph: 
given the spherical harmonic modes of $\varphi_1$ \emph{up to some maximum 
$l=\lmax$}, how can we accurately compute the modes of $S$?

\begin{figure}[t]
\begin{center}
\includegraphics[width=.95\columnwidth]{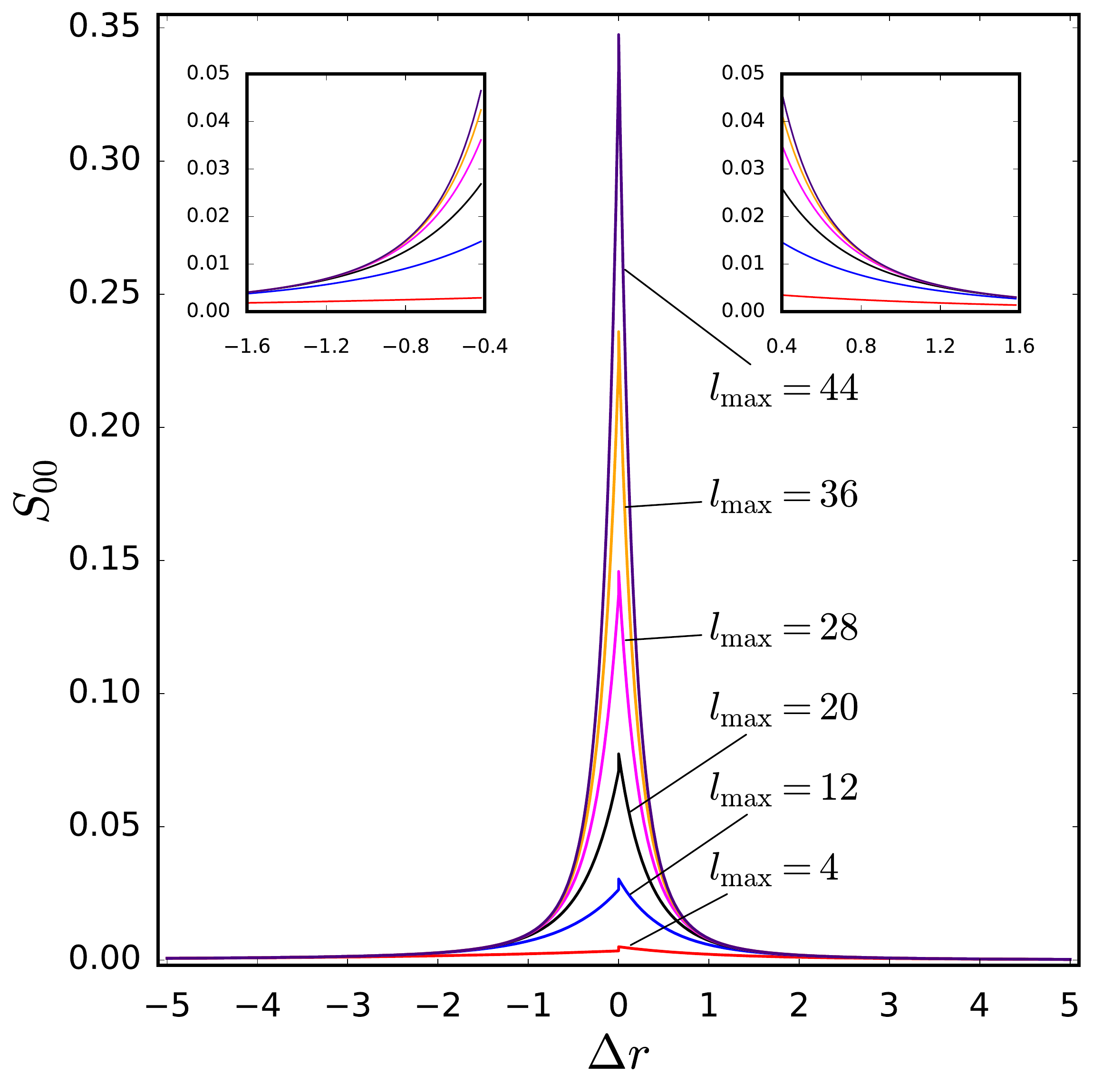}
\end{center}
\caption{\label{Fig:slow convergence} The source mode $S_{00}[\varphi^{\rm 
ret},\varphi^{\rm ret}]$ as a function of $\Delta r:=r-r_0$, with an orbital 
radius $r_0=10$, as computed from the mode-coupling formula~\eqref{Slm}. To 
assess the convergence of the sum in Eq.~\eqref{Slm}, we truncate the 
first-order field modes $\varphi^{\rm}_{lm}$ at a maximum $l$ value $\lmax$, 
and we display the behavior of $S_{00}$ for various values of $\lmax$. The 
insets show that far from the particle, the sum converges rapidly with $\lmax$. 
However, near the particle there is no evidence of numerical convergence.}
\end{figure}
 
Our answer to this question is to utilize a 4D approximation to the 
point-particle singularity. As is well known, the retarded field of a point 
particle can be split into two pieces as $\varphi^{\rm 
ret}=\varphi^S+\varphi^R$~\cite{Detweiler-Whiting:03}, where $\varphi^S$ is the 
Detweiler-Whiting singular field, which is a particular solution to 
Eq.~\eqref{phi1}, and $\varphi^R$ is the corresponding regular field, which is 
a smooth  solution to $\Box \phi^R=0$. The slow falloff of $\varphi^{\rm 
ret}_{lm}$ with $l$ is entirely isolated in the modes of the singular field, 
$\varphi^S_{lm}$; because $\varphi^R$ is smooth, its modes $\varphi^R_{lm}$ 
have a uniform exponential falloff with $l$. Generally, there is no way to 
obtain a closed-form expression for $\varphi^S$, but we \emph{can} easily 
obtain a local expansion of $\varphi^S$ in powers of distance from the particle. 
A truncation of that expansion at some finite order provides a puncture, of the 
sort alluded to above, which we denote by $\varphi^\P$; it is given explicitly 
by Eq.~\eqref{phiP} below. It defines a residual field  
$\varphi^\res:=\varphi^{\rm ret}-\varphi^\P$ that approximates $\varphi^R$. We 
make use of all this by writing the source in the suggestively quadratic form 
$S[\varphi,\varphi]$, and in some region near the particle, splitting the field 
into the two pieces $\varphi^{\P}+\varphi^{\res}$. An $lm$ mode of $S$ can then 
be written as
\begin{align}\label{S pieces}
S_{lm} &=  
S_{lm}[\varphi^{\res},\varphi^{\res}]+2S_{lm}[\varphi^{\res},\varphi^{\P}]+ 
S_{lm}[\varphi^{\P},\varphi^{\P}].
\end{align}
The first two terms, $S_{lm}[\varphi^{\res},\varphi^{\res}]$ and 
$S_{lm}[\varphi^{\res},\varphi^{\P}]$, can be computed from the modes of 
$\varphi^\res$ and $\varphi^\P$ using Eq.~\eqref{Slm}; for sufficiently smooth 
$\varphi^{\res}$, the convergence will be sufficiently rapid. The problem of 
slow convergence is then isolated in the third term, 
$S_{lm}[\varphi^{\P},\varphi^{\P}]$. This term cannot be accurately computed 
from the modes of  $\varphi^\P$. However, $S[\varphi^{\P},\varphi^{\P}]$ 
\emph{can} be computed in 4D using the 4D expression for $\varphi^\P$. Its 
modes $S_{lm}[\varphi^{\P},\varphi^{\P}]$ can then be computed directly, 
without utilizing the mode-coupling formula~\eqref{Slm}, simply by integrating 
the 4D expression against a scalar harmonic.

Our strategy is hence summarized as follows:
\begin{enumerate}
\item compute the modes $\varphi^{\P}_{lm}$ by direct integration of the 4D 
expression~\eqref{phiP-regularized}. From the result, and 
Eqs.~\eqref{phi1lm!=0}--\eqref{phi1l0}, compute the modes 
$\varphi^\res_{lm}=\varphi^{\rm ret}_{lm}-\varphi^\P_{lm}$
\item evaluate $S_{lm}[\varphi^{\res},\varphi^{\res}]$ and 
$S_{lm}[\varphi^{\res},\varphi^{\P}]$ using the mode-coupling 
formula~\eqref{Slm}
\item evaluate $S[\varphi^{\P},\varphi^{\P}]$ in 4D, using 
Eq.~\eqref{phiP-regularized}, and obtain its modes 
$S_{lm}[\varphi^{\P},\varphi^{\P}]$ by direct integration
\item combine these results in Eq.~\eqref{S pieces}.
\end{enumerate}
This strategy is to be applied in some region around $r=r_0$; outside that 
region, one may simply use the retarded modes in Eq.~\eqref{Slm} without 
difficulty.

\begin{figure}[t]
\begin{center}
\includegraphics[width=.93\columnwidth]{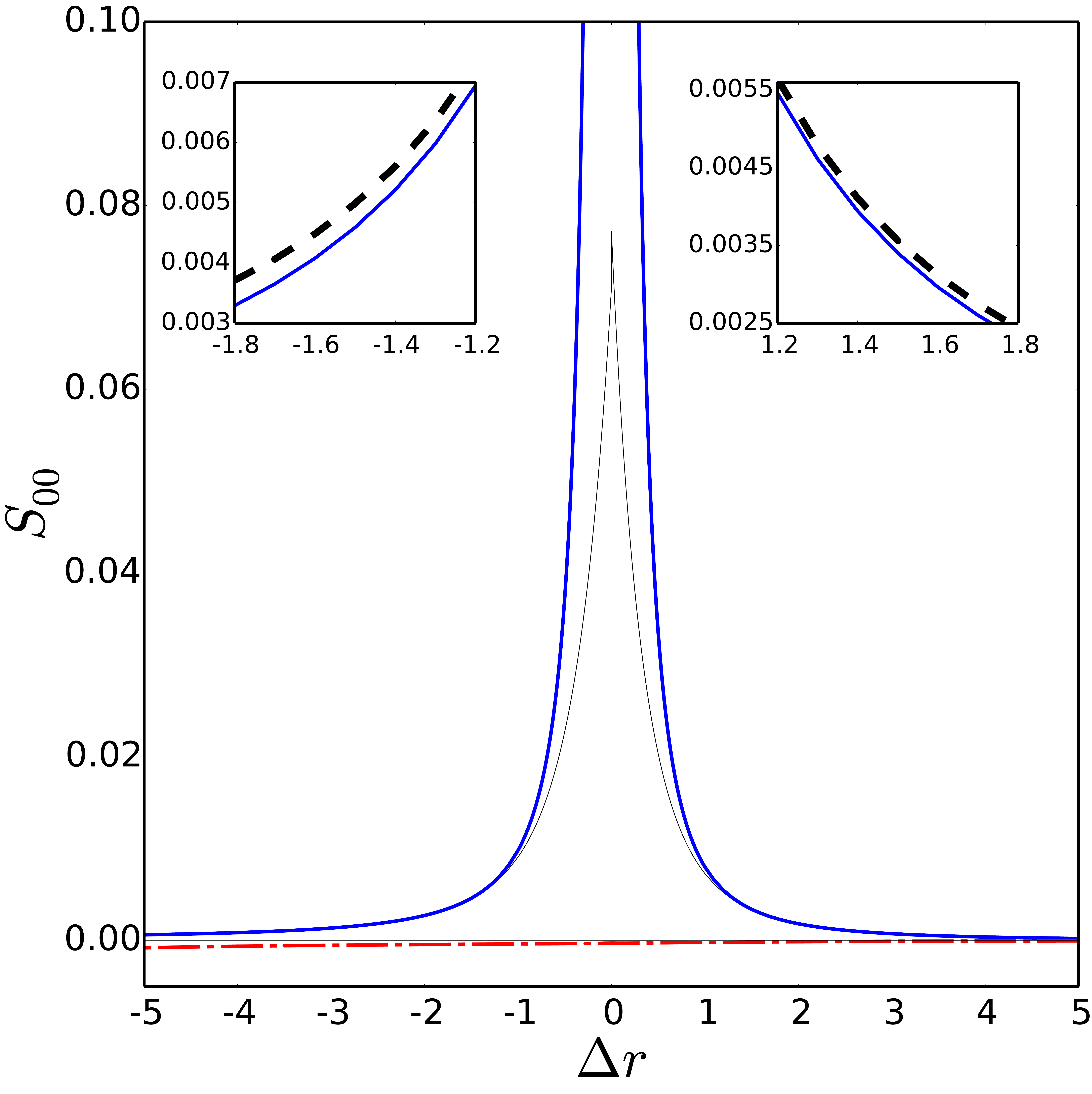}
\end{center}
\caption{\label{Fig:together} The source mode $S_{00}[\varphi^{\rm 
ret},\varphi^{\rm ret}]$ 
and its two contributions as functions of $\Delta r$, as computed with the 
strategy outlined in the text. The dot-dashed red curve shows the contribution 
from 
$S_{00}[\varphi^{\res},\varphi^{\res}]+2S_{00}[\varphi^{\res},\varphi^{\P}]$, 
the dashed black curve shows the contribution from 
$S_{00}[\varphi^{\P},\varphi^{\P}]$, and the thick solid blue curve shows their 
sum $S_{00}[\varphi^{\rm ret},\varphi^{\rm ret}]$, which diverges at $\Delta 
r=0$. On the scale of the main plot, $S_{00}[\varphi^{\P},\varphi^{\P}]$ is 
indistinguishable from $S_{00}[\varphi^{\rm ret},\varphi^{\rm ret}]$; the 
insets show that they differ by a small, but distinguishable amount, which is 
made up by 
$S_{00}[\varphi^{\res},\varphi^{\res}]+2S_{00}[\varphi^{\res},\varphi^{\P}]$. 
For comparison, the thin grey curve displays the result for 
$S_{00}[\varphi^{\rm ret},\varphi^{\rm ret}]$ as computed from the mode-coupling 
formula~\eqref{Slm}, which agrees with the correct result far from the particle 
but differs strongly from it near the particle. All curves were generated with 
$r_0=10$, all four orders in the puncture~\eqref{phiP}, and $\lmax=20$.}\end{figure}

Figure~\ref{Fig:together}  displays a successful implementation of this 
strategy. The true source mode $S_{00}$, as computed via our strategy, is shown 
in thick solid blue. The same mode  $S_{00}$ as computed via mode coupling from 
$\varphi^{\rm ret}_{lm}$, with a finite $\lmax$, is shown in thin solid grey. 
As we can see, the two results agree far from the particle, where the source 
mode as computed via mode coupling has converged. But near the particle, the 
results differ by an arbitrarily large amount; the true source correctly 
diverges at $r=r_0$, due to the singularity in the first-order field, while the 
source computed via mode coupling remains finite due to the truncation at 
finite $\lmax$. 

In the remaining sections, we describe the technical details of our strategy, 
as well as the challenges that arise in implementing it. Section~\ref{fields 
and coupling} summarizes the various relevant fields---retarded and advanced, 
singular and regular, puncture and residual. Section~\ref{source} derives the 
coupling formula that expresses a second-order source mode $S_{lm}$ as a sum 
over first-order field modes. Section~\ref{SR and RR} details the computation 
of $S_{lm}[\varphi^{\res},\varphi^{\res}]$ and 
$S_{lm}[\varphi^{\res},\varphi^{\P}]$; Sec.~\ref{SS}, the computation of  
$S_{lm}[\varphi^{\P},\varphi^{\P}]$. In Sec.~\ref{Conclusion}, we reiterate the outline of our 
strategy as it applies to the gravitational case; the successful application to gravity, recently reported in Ref.~\cite{Pound:16b}, will be detailed in a future paper.

To avoid repetition, we state in advance that all plots are for a particle at radius $r_0=10$. 

\section{First-order fields}\label{fields and coupling}

\subsection{Retarded and advanced solutions}
To begin, we work in spherical polar coordinates $(t,r,\theta^A)$, where 
$\theta^A:=(\theta,\phi)$. We place the particle on the equatorial circular 
orbit $x_p^\mu(t)=(t,r_0,\pi/2,\Omega t)$ with normalized four-velocity 
$u^\mu=(1-r_0^2\Omega^2)^{-1/2}(1,0,0,\Omega)$, and we adopt a Keplerian 
frequency $\Omega=\sqrt{1/r_0^3}$. The point source~\eqref{rho} can then be 
expanded in spherical and frequency harmonics by rewriting it as
\beq\label{rho-lmw}
\varrho= \frac{\delta(r-r_p)}{r^2u^t}\sum_{ l m}Y^*_{ l m}(\theta^A_p)Y_{ l 
m}(\theta^A)
\eeq
and using $Y^*_{ l m}(\theta^A_p)=e^{-im\Omega t}Y_{lm}(\pi/2,0)$. Here 
$u^t=\frac{dt}{d\tau}=(1-r_0^2\Omega^2)^{-1/2}$. 

Most of the fields we are interested in can be constructed by integrating this 
source against a Green's function. The retarded and advanced Green's functions 
satisfying $\Box G(x,x') = -4\pi \delta^4(x-x')$ are given by
\beq
G^{\rm ret/adv}(x,x') =  
\frac{\delta(t-t'\mp|\vec{x}-\vec{x}'|)}{|\vec{x}-\vec{x}'|},
\eeq
where $\vec{x}$ is a Cartesian three-vector. The Fourier transforms, $G^{\rm 
ret/adv}_\omega = \int e^{i\omega (t-t')}G^{\rm ret/adv}(x,x')dt$, are
\beq
G^{\rm ret/adv}_\omega =  \frac{e^{\pm 
i\omega|\vec{x}-\vec{x}'|}}{|\vec{x}-\vec{x}'|},
\eeq
which can be expanded in spherical harmonics as
\beq
G^{\rm ret/adv}_\omega = \mp i\sum_{lm}\omega j_l(\omega r_<)h^{(1,2)}_l(\omega 
r_>)Y^*_{lm}(\theta^{A'}) Y_{lm}(\theta^A).
\eeq
Here the upper sign and $h^{(1)}_{l}$ correspond to the retarded solution, and 
the lower sign and $h^{(2)}_{l}$ to the advanced. $h^{(1)}_{l}$ and 
$h^{(2)}_{l}$ are the spherical Hankel functions of the first and second kind, 
$j_l$ is the spherical Bessel function of the first kind, and when used in the 
Green's function, $r_{\lessgtr}:={\rm min/max}(r,r')$. In the static limit 
$\omega\to0$, the retarded and advanced Green's functions both reduce to
\beq
G^{\rm ret/adv} = \frac{1}{|\vec{x}-\vec{x}'|} = \sum_{ l 
m}\frac{1}{2l+1}\frac{r_<^l}{r_>^{l+1}}Y^*_{ l m}(\theta^{A'})Y_{ l 
m}(\theta^A).
\eeq

Integrating against these Green's functions, we find the retarded and advanced 
solutions
\beq\label{phi-lmw}
\varphi^{\rm ret/adv} = \sum_{lm}\varphi^{\rm ret/adv}_{l m}(r)e^{-im\Omega 
t}Y_{lm}(\theta^A),
\eeq
where
\begin{align}\label{phi1lm!=0}
\varphi^{\rm ret/adv}_{lm} &= \pm\frac{4\pi i}{u^t}N_{l m} m \Omega j_l(m\Omega 
r_<)h^{(1,2)}_{l}(m\Omega r_>)
\end{align}
for $m\neq0$, and
\beq\label{phi1l0}
\varphi^{\rm ret/adv}_{l 0} = 
\frac{4\pi}{u^t}\frac{N_{l0}}{2l+1}\frac{r_<^l}{r_>^{l+1}}
\eeq
for $m=0$. Here $N_{l m}:=Y_{lm}(\pi/2,0)$, and we have reverted to the 
previous notation $r_{\lessgtr}:={\rm min/max}(r,r_0)$.

As discussed in the introduction, the large-$l$ behavior of these fields is the 
source of the infinite-coupling problem. Noting that $N_{l0}\sim l^0$, we see 
that the stationary modes in Eq.~\eqref{phi1l0} behave as $\varphi_{l 0}\sim 
\frac{1}{l}\frac{r_<^l}{r_>^{l+1}}$. Hence, $\varphi_{l0}$ decays exponentially 
with $l$ at points far from $r=r_0$, still exponentially but more slowly at 
points close to $r=r_0$, and as $l^{-1}$ at $r=r_0$. The oscillatory, $m\neq0$ 
modes  exhibit similar behavior, although it is not obvious from 
Eq.~\eqref{phi1lm!=0}. After summing $\varphi_{lm}Y_{lm}$ over $m$, the 
large-$l$ behavior becomes $\sim l^0$ on the particle, with an exponential but 
arbitrarily weak suppression at points slightly off the particle. The 
quantitative consequences of this, already displayed in Fig.~\ref{Fig:slow 
convergence}, will be spelled out in later sections. 

\subsection{Singular and regular fields}
In flat space, the Detweiler-Whiting singular field is simply 
$\varphi^S:=\frac{1}{2}(\varphi^{\rm ret}+\varphi^{\rm adv})$. Its 
four-dimensional form can be written as 
\beq\label{phiS-4D}
\varphi^S = \frac{1}{2}\int [G^{\rm ret}(x,x')+G^{\rm 
adv}(x,x')]\varrho(x')d^4x'.
\eeq
Its modes are more easily found directly from Eqs.~\eqref{phi1lm!=0} and 
\eqref{phi1l0}. For $m\neq0$,
\beq
\varphi^S_{lm} =\frac{4\pi}{u^t}N_{l m}  m \Omega j_l(m\Omega r_<)y_{l}(m\Omega 
r_>),
\eeq
where $y_l$ is the spherical Bessel function of the second kind. For $m=0$, 
$\varphi^S_{l0}=\varphi^{\rm ret/adv}_{l0}$.

Correspondingly, in flat space the regular field is $\varphi^R=\varphi^{\rm 
ret}-\varphi^\S=\frac{1}{2}(\varphi^{\rm ret}-\varphi^{\rm adv})$. Its 
four-dimensional form can be written as an integral analogous 
to~\eqref{phiS-4D}. Its modes can be found straightforwardly from 
Eqs.~\eqref{phi1lm!=0} and \eqref{phi1l0}. For $m\neq0$,
\beq
\varphi^R_{lm} =\frac{4\pi i}{u^t}N_{l m}  m \Omega j_l(m\Omega 
r_<)j_{l}(m\Omega r_>), 
\eeq
and for $m=0$, $\varphi^R_{l0}=0$.

\subsection{Puncture and residual fields}
The puncture field $\varphi^\P$ is obtained in 4D by performing a local 
expansion of the integral representation~\eqref{phiS-4D} of the singular field. 
That procedure is common in the literature, and so we do not belabor it here; 
instead we refer the reader to, e.g., Ref.~\cite{Heffernan-etal:12} for 
details, and give here only the main results. Letting $\lambda:=1$ count powers
of distance from the particle, the covariant expansion of the
flat-space puncture to fourth-from-leading order in distance is
\begin{widetext}
\begin{align}
\label{eq:cov-puncture}
\varphi^{\S}(x; x_p) =&{} \frac{1}{\bar{s}}
 + \frac{\sigma _a \left(\bar{s}^2-\bar{r}^2\right)}{2 \bar{s}^3}
 + \frac{a^2 \bar{s}^2 \left(\bar{r}^4-6 \bar{r}^2 \bar{s}^2-3 \bar{s}^4\right)+9 \sigma _a^2 \left(\bar{r}^2-\bar{s}^2\right)^2-4 \bar{r} \bar{s}^2 \sigma _{\dot{a}} \left(\bar{r}^2-3 \bar{s}^2\right)}{24 \bar{s}^5}
\nonumber \\
 &+\frac{1}{48 \bar{s}^7}\Big[
  2 \bar{r} \bar{s}^4 a^\alpha \dot{a}_{\alpha} \left(\bar{r}^4-10 \bar{r}^2 \bar{s}^2-15 \bar{s}^4\right)
  -3 a^2 \bar{s}^2 \sigma _a \left(\bar{r}^6-5 \bar{r}^4 \bar{s}^2+15 \bar{r}^2 \bar{s}^4+5 \bar{s}^6\right)
\nonumber \\
 & \quad + 4 \sigma _a \sigma _{\dot{a}} \bar{r} \bar{s}^2 (3 \bar{r}^4-10 \bar{r}^2 \bar{s}^2 +15 \bar{s}^4)
  - 15 \sigma _a^3 (\bar{r}^2-\bar{s}^2)^3
  - 2 \sigma _{\ddot{a}} \bar{s}^4 ( \bar{r}^4 -6 \bar{r}^2 \bar{s}^2 - 3 \bar{s}^4)
 \Big] + \mathcal{O}(\lambda^3).
\end{align}
\end{widetext}
where the terms are $\mathcal{O}(\lambda^{-1})$, $\mathcal{O}(\lambda^0)$,
$\mathcal{O}(\lambda^1)$ and $\mathcal{O}(\lambda^2)$, respectively.
Here, we follow the notation of Ref.~\cite{Heffernan-etal:12}: we make use of
the compact notation of Haas and Poisson~\cite{Haas-Poisson:06},
$\sigma_X := \sigma_{\alpha} X^\alpha$ for any vector $X^\alpha$; the bi-scalar
$\sigma(x,x_p)$ is the Synge world function, equal to one half of the
squared geodesic distance between $x$ and $x_p$, and $\sigma_\alpha:=\frac{\partial\sigma}{\partial x_p^\alpha}$; the vectors
$a^\alpha := u^\beta \nabla_\beta u^\alpha$,
$\dot{a}^\alpha := u^\beta \nabla_\beta a^\alpha$ and
$\ddot{a}^\alpha := u^\beta \nabla_\beta \dot{a}^\alpha$ are the acceleration
and its first and second derivatives, respectively; and the quantities
$\bar{r} := \sigma_{\alpha} u^\alpha$ and
$\bar{s} := \sqrt{(g^{\alpha\beta}+u^\alpha u^\beta) \sigma_{\alpha} \sigma_{\beta}}$
are projected components of the geodesic distance from the field point to the
reference point $x_p$ on the worldline. In our case, $g_{\alpha\beta}$ is the metric of flat spacetime and $\nabla_\alpha$ is the covariant derivative compatible with it.

To facilitate the computation of spherical harmonic modes, it is customary to 
express the field in a rotated coordinate system in which the particle is 
momentarily at the north pole. We label the angles in this system 
$\alpha^{A'}:=(\alpha,\beta)$, such that at a given instant $t$, the particle 
sits at $\alpha=0$. More details can be found in the Appendix. As we describe 
there and in later sections, in our calculations this rotation introduces new 
complications and loses some of its traditional advantages. Nevertheless, its 
benefits outweigh its drawbacks.

In terms of the rotated angles $\alpha^{A'}$, a puncture satisfying
$\varphi^\P = \varphi^S +\O(\lambda^3)$ can be obtained from a coordinate
expansion of Eq.~\eqref{eq:cov-puncture}. For the circular orbits we are
interested in here, this is given explicitly by
\beq\label{phiP}
\varphi^{\P} = \lambda^{-1}\varphi^{\P}_{(-1)} + 
\lambda^{0}\varphi^{\P}_{(0)}+\lambda\varphi^{\P}_{(1)}+\lambda^2\varphi^{\P}_{
(2)},
\eeq
where
\begin{subequations}\allowdisplaybreaks\label{eq:singular-field}
\begin{align}
\varphi^{\P}_{(-1)} &= \frac{1}{\rho} ,\\
\varphi^{\P}_{(0)} &= -\frac{\Delta r  }{2 r_0 \rho\chi } (1- 2 v^2 s^2) 
\nonumber\\
							&\quad + \frac{\Delta 
r^3}{2r_0\chi_0\chi\rho^3 } (1 - 2 v^2  s^2  +v^4  s^2 ),\\
\varphi^{\P}_{(1)} &= \frac{3 \Delta r^6}{8 r_0^2 \rho^5 \chi_0^2\chi ^2} 
\left(1 - 2v^2  s^2  + v^4  s^2 \right)^2 \nonumber\\
							&\quad + \frac{\rho 
v^2}{24r_0^2 \chi_0^2\chi ^2} [3 v^6  s^2  - 3 (1 +  s^2 ) - 3 v^2 (2 - 7  s^2 
) \nonumber\\
							&\quad + v^4 (1 - 5  
s^2  - 8  s^4 )]  +\frac{\Delta r^2}{24 r_0^2 \rho \chi_0^2\chi ^2} [9 
\nonumber\\
							&\quad - 18 v^2 (1 +  
s^2 ) - 6 v^8  s^2  (1 - 4  s^2 )  \nonumber\\
							&\quad + 3v^4 (5 + 8  
s^2  + 8  s^4 ) - 2 v^6 (1 + 4  s^2  + 22  s^4 )] \nonumber\\
							&\quad + \frac{\Delta 
r^4}{24 r_0^2 \rho^3 \chi_0^2\chi ^2} [-18 + 3 v^8  s^2  (1 - 9  s^2 )  
\nonumber\\
							&\quad + 3 v^2 (7 + 19  
s^2 ) - 3 v^4 (1 + 21  s^2  + 20  s^4 )  \nonumber\\
							&\quad + v^6 (1 +  s^2  
+ 88  s^4 )],\\
\varphi^{\P}_{(2)} &= \frac{5 \Delta r^9}{16 r_0^3 \rho^7 \chi_0^3\chi ^3} (1 
- 2 v^2  s^2  + v^4  s^2)^3 \nonumber\\
							&\quad -  \frac{\Delta 
r \rho v^2}{48 r_0^3 \chi_0^3\chi ^3}  [6 v^{10}  s^4  + 3 (1 +  s^2 ) 
\nonumber\\
							&\quad + v^8  s^2  (7 - 
8  s^2  - 32  s^4 ) + 3 v^2 (11 - 14  s^2  + 2  s^4 ) \nonumber\\
							&\quad + v^4 (13 - 62  
s^2  + 16  s^4 ) \nonumber\\
							&\quad -  v^6 (1 + 50  
s^2  - 124  s^4  + 16  s^6)] \nonumber\\
							&\quad -  \frac{\Delta 
r^7}{16 r_0^3 \rho^5 \chi_0^3\chi ^3} [15 - 3 v^{12}  s^4  (1 - 7  s^2 ) 
\nonumber\\
							&\quad - 3 v^2 (6 + 25  
s^2 ) + 3 v^4 (1 + 33  s^2  + 46  s^4 ) \nonumber\\
							&\quad -  v^{10}  s^2  
(1 - 8  s^2  + 112  s^4 ) \nonumber\\
							&\quad + v^8  s^2  (2 + 
65  s^2  + 188  s^4 ) \nonumber\\
							&\quad - v^6 (1 + 22  
s^2  + 211  s^4  + 96  s^6)]  \nonumber\\
							&\quad -  \frac{\Delta 
r^3}{48 r_0^3 \rho \chi_0^3\chi ^3} [15 - 3 v^{12}  s^4  (7 - 16  s^2 ) 
\nonumber\\
							&\quad - 3 v^2 (16 + 17 
 s^2 ) - v^{10}  s^2  (17 - 13  s^2  + 128  s^4 ) \nonumber\\
							&\quad + 3v^4 (11 + 61  
s^2  + 14  s^4 ) \nonumber\\
							&\quad -  v^6 (26 + 158 
 s^2  + 125  s^4  + 48 s^6) \nonumber\\
							&\quad + v^8 (2 + 115  
s^2  + 19  s^4  + 152  s^6)]  \nonumber\\
							&\quad + \frac{\Delta 
r^5}{48 r_0^3 \rho^3 \chi_0^3\chi ^3} [45 - 6 v^{12}  s^4  (4 - 15  s^2 ) 
\nonumber\\
							&\quad - 3 v^2 (33 + 61 
 s^2 ) - v^{10}  s^2  (13 - 47  s^2  + 400  s^4 ) \nonumber\\
							&\quad + 3 v^4 (23 + 
131  s^2  + 94  s^4 ) \nonumber\\
							&\quad - 2 v^6 (5 + 134 
 s^2  + 281  s^4  + 108 s^6) \nonumber\\
							&\quad + v^8 (1 + 53  
s^2  + 275  s^4  + 520  s^6)].
\end{align}
\end{subequations}
Here $v^2:=r_0^2\Omega^2$, $s:=\sin\beta$, $\chi:= 1 -  v^2  s^2$, $\chi_0:= 1 
-  v^2=1/(u^t)^2$, and
\begin{equation}
\label{eq:rho}
 \rho:=\Big[\frac{2 r_0^2 \chi}{\chi_0} (\delta^2+1-\cos\alpha)\Big]^{1/2},
\end{equation}
with 
$\delta^2:=\frac{\chi_0\Delta r^2}{2r_0^2\chi}$. Note that the only dependence 
of the singular field on $\alpha$ appears through $\rho$, while $\beta$ appears 
through $\rho$,  $\chi$, and the explicit powers of $s$. Also note that the 
above expression for $\varphi^\P(\alpha^{A'})$ is valid only at the instant 
when the particle is at the north pole of the rotated coordinate system.

Given this choice of puncture field, the residual field is defined implicitly 
by $\varphi^\res:=\varphi^{\rm ret}-\varphi^\P$. Since we do not have a 
closed-form expression for $\varphi^{\rm ret}$, we cannot write an exact result 
for $\varphi^\res$ in 4D. However, we can compute its modes from those of 
$\varphi^{\rm ret}$ and $\varphi^\P$ using $\varphi^\res_{lm}=\varphi^{\rm 
ret}_{lm}-\varphi^\P_{lm}$.

Before proceeding, note that in Eq.~\eqref{phiP}, we have kept the first four 
orders from the local expansion of $\varphi^\S$. We refer to this as a 
fourth-order puncture; if in a particular calculation we include only the first 
three of them, we refer to it as a third-order puncture, and so on. The higher 
the order of the puncture, the smoother the residual field, and hence the more 
rapid the falloff of $\varphi^\res_{lm}$ with $l$. In the following sections we 
will explore how our strategy of computing $S$ is impacted by this, and we 
shall find that the puncture must be of at least third order for our strategy 
to succeed.

\section{Second-order source}\label{source}
We are now interested in how the modes of the fields are coupled in the source 
$S=t^{\mu\nu}\partial_\mu\varphi_1\partial_\nu\varphi_1$. For later use, we 
derive the mode-coupling formula in both $\theta^A$ and $\alpha^{A'}$ 
coordinates. The method of derivation, and the end result in $\theta^A$ 
coordinates, was previously presented in Ref.~\cite{Pound:15c}, and so we 
omit some details  here. 

\subsection{In $\theta^A$ coordinates}
Written as a bilinear functional, $S$ is given more explicitly by
\begin{widetext}
\begin{align}
S[\varphi^{(1)},\varphi^{(2)}] &= \partial_t \varphi^{(1)} 
\partial_t\varphi^{(2)}+ \partial_r \varphi^{(1)}\partial_r \varphi^{(2)} 										
 + 
\frac{1}{r^2}\Omega^{AB}\partial_A\varphi^{(1)}\partial_B\varphi^{(2)},\label{unrotated S}
\end{align}
where $\varphi^{(1)}$ and $\varphi^{(2)}$ are any two differentiable fields, 
$\Omega_{AB}={\rm diag}(1,\sin^2\theta)$ is the metric of the unit sphere and 
$\Omega^{AB}$ is its inverse. Substituting 
$\varphi^{(n)}=\sum_{lm}\varphi^{(n)}_{lm}(r)e^{-im\Omega t}Y_{lm}$, we get 
\begin{align}
S &= \sum_{\substack{l_1m_1\\l_2m_2}}e^{-i(m_1+m_2)\Omega t}\Big[\big( 
\partial_r\varphi^{(1)}_{l_1m_1}\partial_r\varphi^{(2)}_{l_2m_2}
		 -m_1m_2\Omega^2 
\varphi^{(1)}_{l_1m_1}\varphi^{(2)}_{l_2m_2}\big)Y_{l_1m_1}Y_{l_2m_2} 
		 + 
\frac{1}{r^2}\varphi^{(1)}_{l_1m_1}\varphi^{(2)}_{l_2m_2}\partial^AY_{l_1m_1}
\partial_AY_{l_2m_2}\Big],\label{Ssextuplesum}
\end{align}
where indices are raised with $\Omega^{AB}$. 

To obtain the spherical-harmonic coefficient of Eq.~\eqref{Ssextuplesum}, we first rewrite $\partial_AY_{lm}$ in terms of spin-weighted harmonics ${}_sY_{lm}$ as 
\begin{equation}\label{DYtosY}
\partial_A Y^{\ell m} = \frac{1}{2}\sqrt{\ell(\ell+1)}\left({}_{-1}Y^{\ell m}m_A-{}_1Y^{\ell m}m^*_A\right),
\end{equation}
where $m^A:=\left(1,\frac{i}{\sin\theta}\right)$ and its complex conjugate $m^{*A}$ form a null basis on the unit sphere. This allows us to compute $S_{lm}$, which is an integral against $Y^*_{lm}={}_0Y^*_{lm}$, by appealing to the general formula 
\begin{equation}\label{Cdef}
\oint {}_{s}Y^{l m*}{}_{s_1}Y^{l_1 m_1}{}_{s_2}Y^{l_2 m_2}d\Omega = C^{l m 
s}_{l_1m_1s_1l_2m_2s_2}, 
\end{equation}
where $d\Omega=\sin\theta\, d\theta\, d\phi$ and for $s=s_1+s_2$, 
\begin{align}\label{coupling}
C^{l m s}_{l_1m_1s_1l_2m_2s_2} &= 
(-1)^{m+s}\sqrt{\frac{(2l+1)(2l_1+1)(2l_2+1)}{4\pi}}
					\begin{pmatrix}l & l_1 & l_2 \\ s & 
-s_1 & -s_2\end{pmatrix}
					\begin{pmatrix}l & l_1 & l_2 \\ -m & 
m_1 & m_2\end{pmatrix}\!.\!
\end{align}
Here the arrays are $3j$ symbols. If $s=s_1=s_2=0$, Eq.~\eqref{Cdef} reduces to 
the standard formula for the integral of three ordinary spherical harmonics. We 
refer the reader to Ref.~\cite{Pound:15c} for more details.

After  using Eq.~\eqref{DYtosY}, $m^Am_A=0$, $m^Am_A^*=2$, and Eq.~\eqref{Cdef}, we find that Eq.~\eqref{Ssextuplesum} can be written as $S=\sum_{l 
m}S_{l m}(r)e^{-im\Omega t}Y_{l m}$, with modes given by
\begin{align}\label{Slm}
S_{l m}[\varphi^{(1)},\varphi^{(2)}] &=  
\sum_{\substack{l_1m_1\\l_2m_2}}\!\bigg[
			C^{l m 0}_{l_1m_10l_2m_20}\left(\partial_r 
\varphi^{(1)}_{l_1m_1}\partial_r \varphi^{(2)}_{l_2m_2} 
				-m_1m_2\Omega^2 
\varphi^{(1)}_{l_1m_1}\varphi^{(2)}_{l_2m_2}\right)\nonumber\\
&\quad	-\frac{1}{2r^2}\sqrt{l_1(l_1+1)l_2(l_2+1)}C^{l m 
0}_{l_1m_1-1l_2m_21}\left(\varphi^{(1)}_{l_1m_1}\varphi^{(2)}_{l_2m_2}+\varphi^{
(2)}_{l_1m_1}\varphi^{(1)}_{l_2m_2}\right) \bigg]\!.
\end{align}
\end{widetext}
We have used the freedom to relabel $l_1m_1\leftrightarrow l_2m_2$ and the
symmetry $C^{l m s}_{l_1m_1s_1l_2m_2s_2}=C^{l m s}_{l_2m_2s_2l_1m_1s_1}$ to 
slightly simplify this result. We note that the range of the sum is restricted by the $3j$ symbols in $C^{l m 
s}_{l_1m_1s_1l_2m_2s_2}$, which enforce (i) $m=m_1+m_2$ and (ii) the triangle 
inequality $|l_1-l_2|\leq l\leq l_1+l_2$. The first of these  restrictions has 
been used to replace $e^{-i(m_1+m_2)\Omega t}$ with $e^{-im\Omega t}$, and it 
can be further used to eliminate the sum over $m_2$.

In our toy model, Eq.~\eqref{Slm} plays the role of Eq.~\eqref{ddGilm} from the 
gravitational case. When we only have access to a finite number of modes 
$\varphi^{(n)}_{lm}$ up to $l=\lmax$, then the sum is truncated: explicitly, it 
becomes the partial sum
\beq\label{Slmax}
S^\lmax_{lm} := \sum_{l_1=0}^\lmax\sum_{l_2=0}^\lmax 
\sum_{m_1=-l_1}^{l_1}S^{l_1m_1l_2,m-m_1}_{lm},
\eeq
where we have eliminated the sum over $m_2$, and for brevity we have suppressed 
the functional arguments and defined $S^{l_1m_1l_2m_2}_{lm}$ as the summand in 
Eq.~\eqref{Slm}. By appealing to the triangle inequality, we could write the second sum even more explicitly as 
$\sum_{l_2=|l-l_1|}^{{\rm min}(\lmax,l+l_1)}$. 

The slow convergence of the limit $S^\lmax_{lm}\to S_{lm}$ was illustrated in 
Fig.~\ref{Fig:slow convergence}. Its behavior will be more carefully analyzed 
in the following sections.

\subsection{In $\alpha^{A'}$ coordinates}

Although Eq.~\eqref{Slm} is the mode-coupling formula that we will utilize in 
explicit computations, we will also make use of the analogous formula in the 
rotated coordinates $\alpha^{A'}$. Deriving that result additionally provides an 
opportunity to introduce the 4D form of $S$  in these coordinates, which will 
be essential in Sec.~\ref{SS}.

Obtaining the source in the rotated coordinates involves a new subtlety: the 4D 
expression for $S$ involves $t$ derivatives, while our expression \eqref{phiP} 
for $\varphi^\P(\alpha^{A'})$ is intended to only be instantaneously valid at 
the instant when the particle is at the north pole of the rotated coordinate 
system. We discuss this subtlety in Appendix~\ref{rotations}. In brief, we may 
treat the coordinates $\alpha^{A'}$ as themselves dependent on $t$, and 
appropriately account for that time dependence when acting with $t$ 
derivatives. The 4D expression for $S$ is then given by Eq.~\eqref{rotated S}, 
which we reproduce here for convenience:
\begin{align}
S[\varphi^{(1)},\varphi^{(2)}] &= \dot\alpha^{A'}\partial_{A'} \varphi^{(1)} 
\dot\alpha^{A'}\partial_{A'}\varphi^{(2)} + \partial_r \varphi^{(1)}\partial_r 
\varphi^{(2)} \nonumber\\										
	&\quad + 
\frac{1}{r^2}\Omega^{A'B'}\partial_{A'}\varphi^{(1)}\partial_{B'}\varphi^{(1)},
\label{rotated S body}
\end{align}
where $\Omega^{A'B'}={\rm diag}(1,\csc^2\alpha)$ is the inverse metric on the 
unit sphere in the rotated coordinates, and the time derivatives in 
Eq.~\eqref{unrotated S} now manifest in the quantity $\dot\alpha^{A'} = 
\Omega(-\cos\beta,\cot\alpha\sin\beta)$.

The modes of the source in the rotated coordinates are given by
\beq\label{Slm' modes}
S_{lm'} = \oint S(\alpha^{A'})Y^*_{lm'}(\alpha^{A'})d\Omega'.
\eeq 
We will consistently use $m'$ to denote the azimuthal number in the rotated 
coordinates; because $l$ is invariant under rotations, it is the same in both 
sets of coordinates. 

In Sec.~\ref{SS} we will evaluate the integral \eqref{Slm' modes} for 
$S[\varphi^{\P},\varphi^{\P}]$ \emph{without first decomposing 
$\varphi^{\P}$ into modes}. But generically, if we expand each $\varphi^{(n)}$ 
as $\sum_{lm'}\varphi^{(n)}_{lm'}Y_{lm'}$, then we can evaluate the integral 
analytically in the same way as we did for $S_{lm}$. This is made possible by 
first writing $\dot{\alpha}^{A'}$ in terms of spin-weight $\pm 1$ harmonics as
\beq
\dot\alpha^{A'}= \sqrt{\frac{\pi}{3}}\Omega \big[({}_{-1}Y_{11}+{}_{-1}Y_{1,-1})m^{A'}+({}_1Y_{11}+{}_1Y_{1,-1})m^{*A'}\big].
\eeq
Next, we use Eq.~\eqref{DYtosY}, which is covariant on the unit sphere and hence also applies in $\alpha^{A'}$ coordinates. Combining these results, invoking Eqs.~\eqref{Cdef}-\eqref{coupling}, and using the properties of the 3$j$ symbols to simplify, we find
\begin{align}\label{alphadotdphi}
\dot\alpha^{A'}\partial_{A'}\varphi &= \frac{\Omega}{2}\sum_{lm'}(\mu^-_{lm'}\varphi_{l,m'+1}-\mu^+_{lm'}\varphi_{l,m'-1})Y_{lm'},\!
\end{align}
where $\mu_{lm'}^\pm:=\sqrt{(l\pm m')(l\mp m'+1)}$.

Substituting Eq.~\eqref{alphadotdphi} into Eq.~\eqref{rotated S body} and following the same 
procedure as in the previous section, we find
\begin{widetext}
\begin{align}\label{Slm'}
S_{l m'} &=  \sum_{\substack{l_1m'_1\\l_2m'_2}}\!\bigg\{
			C^{l m' 0}_{l_1m'_10l_2m'_20}\big[\partial_r \varphi^{(1)}_{l_1m'_1}\partial_r \varphi^{(2)}_{l_2m'_2} 
			+\tfrac{1}{4}\Omega^2(\mu^-_1\varphi^{(1)}_{l_1,m'_1+1}-\mu^+_1\varphi_{l_1,m_1'-1})(\mu^-_2\varphi_{l_2,m'_2+1}-\mu^+_2\varphi_{l_2,m'_2-1})\big]\nonumber\\
		 	&\quad -\frac{1}{2r^2}\sqrt{l_1(l_1+1)l_2(l_2+1)}C^{l m' 0}_{l_1m'_1-1l_2m'_21}\left(\varphi^{(1)}_{l_1m'_1}\varphi^{(2)}_{l_2m'_2}+\varphi^{(2)}_{l_1m'_1}\varphi^{(1)}_{l_2m'_2}\right)\!\!\bigg\},
\end{align}
\end{widetext}
where $\mu_i^\pm:=\mu^\pm_{l_im'_i}$. Note that unlike Eq.~\eqref{Slm}, which 
gave the coefficient in $\sum_{lm}S_{lm}(r)e^{-im\Omega t}Y_{lm}(\theta^A)$, 
Eq.~\eqref{Slm'} gives the coefficient in 
$\sum_{lm}S_{lm'}(r)Y_{lm}(\alpha^{A'})$, with no phase factor; the time 
dependence is entirely contained in the $\alpha^{A'}$ dependence.

\section{Computing $S_{lm}[\varphi^{\res},\varphi^{\res}]$ and 
$S_{lm}[\varphi^{\res},\varphi^{\P}]$}\label{SR and RR}
Following the strategy outlined in the introduction, we now compute 
$S_{lm}[\varphi^{\res},\varphi^{\res}]$ and 
$S_{lm}[\varphi^{\res},\varphi^{\P}]$ from the modes of $\varphi^{\res}$ and 
$\varphi^{\P}$ using the mode-coupling formula~\eqref{Slm}. In Sec.~\ref{SS} we 
will then complete our strategy by computing 
$S_{lm}[\varphi^{\P},\varphi^{\P}]$ from the 4D expression for $\varphi^{\P}$.

\subsection{Outline of strategy}\label{SR strategy}
As input for $S_{lm}[\varphi^{\res},\varphi^{\res}]$ and 
$S_{lm}[\varphi^{\res},\varphi^{\P}]$ in Eq.~\eqref{Slm}, we require the modes 
$\varphi^\P_{lm}$. We begin by computing the modes 
\beq\label{lm' modes}
\varphi^\P_{lm'} = \oint \varphi^\P(\alpha^{A'})Y^*_{lm'}(\alpha^{A'})d\Omega'
\eeq 
in the rotated coordinates $\alpha^{A'}$. The modes in the unrotated 
coordinates $\theta^A$ are then retrieved using
\beq\label{WignerD-rotation}
\varphi^\P_{lm} =\sum_{m'}\varphi^\P_{lm'}D^l_{mm'}(\pi,\pi/2,\pi/2),
\eeq
where $D^l_{mm'}$ is a Wigner $D$ matrix element. Equation~\eqref{WignerD-rotation} 
yields the modes in a coordinate system in which the particle is on the equator 
at an azimuthal angle $\phi_p=0$. An additional rotation brings it to its 
original position $\phi_p=\Omega t$. The sole effect of that rotation is to 
introduce the phase $e^{-im\Omega t}$: $\varphi_{lm}\to\varphi_{lm}e^{-im\Omega 
t}$.

Given the modes $\varphi^\P_{lm}$, the rest of the procedure is 
straightforward. In summary, it involves four steps:\footnote{We could 
alternatively compute the modes  $S_{lm'}[\varphi^{\res},\varphi^{\res}]$ and 
$S_{lm'}[\varphi^{\res},\varphi^{\P}]$ directly from $\varphi^{\P}_{lm'}$ using 
Eq.~\eqref{Slm'}. $S_{lm}[\varphi^{\res},\varphi^{\res}]$ and 
$S_{lm}[\varphi^{\res},\varphi^{\P}]$ would then be computed using the analogs 
of Eq.~\eqref{WignerD-rotation}.}
\begin{enumerate}
	\item Decompose the puncture field~\eqref{phiP} into $lm'$ modes using 
Eq.~\eqref{lm' modes}.
	\item Use Eq.~\eqref{WignerD-rotation} to obtain the $lm$ modes 
$\varphi^{\P}_{lm}$.
	\item Compute the residual-field modes 
$\varphi^{\res}_{lm}=\varphi^{\rm ret}_{lm}-\varphi^{\P}_{lm}$ 	[with 
$\varphi^{\rm ret}_{lm}$ given in Eqs.~\eqref{phi1lm!=0} and \eqref{phi1l0}].
	\item Use Eq.~\eqref{Slm} to compute 
$S_{lm}[\varphi^{\res},\varphi^{\res}]$ and 
$S_{lm}[\varphi^{\res},\varphi^{\P}]$.
\end{enumerate}
Section~\ref{phiPlm computation} describes the first three steps, and 
Sec.~\ref{RR and RS results} presents and discusses the results of the final 
step.

\subsection{Calculation of $\phi^{\P}_{lm}$}\label{phiPlm computation}
Concretely evaluating the integrals~\eqref{lm' modes} is a nontrivial task. But 
before addressing that topic, we make several prefatory remarks.

First, we note that although integrals like~\eqref{lm' modes} of  local 
expansions like~\eqref{phiP} are common in the literature, in our context they 
introduce a unique challenge. Typically, integrals of this sort appear in 
mode-sum regularization and puncture schemes~\cite{Barack:09,Wardell:15}. In 
those contexts, one's primary goal is to compute the Detweiler-Whiting regular 
field (or some finite number of its derivatives) on the particle's worldline. 
This gives one considerable leeway: If one is interested in computing $n$ 
derivatives of the regular field, for example, then so long as one preserves 
the puncture through order $\lambda^{n}$, one can smoothly deform the integrand 
in Eq.~\eqref{lm' modes}, and one can do so in a different way for each $lm'$ 
mode. Similarly, one can evaluate the integral with a local expansion in the 
limit $\Delta r\to0$, which generally simplifies the integration. And since 
$Y_{lm'}$ vanishes at $\alpha=0$ for $m'\neq0$, one need only evaluate the 
$m'=0$ mode (or in the calculations in Ref.~\cite{Wardell-Warburton:15}, the 
$m'=0,\pm1, \pm2$ modes); traditionally, this restriction to $m'=0$ has been a 
major advantage of using rotated coordinates like $\alpha^{A'}$.

In our calculation, we have none of these luxuries. Because we compute 
$S_{lm}[\varphi^\P,\varphi^\P]$ from the 4D expression for $\varphi^\P$ while 
we compute $S_{lm}[\varphi^\res,\varphi^\P]$ and 
$S_{lm}[\varphi^\res,\varphi^\res]$ from the modes $\varphi^\P_{lm'}$, the modes 
must correspond to an \emph{exact} evaluation of Eq.~\eqref{lm' modes}; 
otherwise, 
$S_{lm}[\varphi^\P,\varphi^\P]+2S_{lm}[\varphi^\res,\varphi^\P]+S_{lm}[
\varphi^\res,\varphi^\res]$ would not be equal to $S_{lm}[\varphi^{\rm 
ret},\varphi^{\rm ret}]$. This means that if we deform the integrand in 
Eq.~\eqref{lm' modes}, then we must make an identical deformation of  the 4D 
expression for $\varphi^\P$. Similarly, any expansion in powers of $\Delta r$ 
would have to be performed for both the $lm'$ modes and the 4D expression; 
because we must evaluate these quantities over a range of $\Delta r$ values, we 
cannot rely on eventually taking the limit $\Delta r\to0$. And finally, we 
cannot limit our computation to $m'=0$; since  we do not evaluate any 
quantities at $\alpha=0$, there is no \emph{a priori} limit to the number of 
$m'$ modes we must compute. (If we only required $S$ on the particle, then we 
would only require the modes $S_{l0'}$, but even these modes depend on all $m'$ 
modes of $\varphi$.)

In brief, we must be exact. We must compute all $lm'$ modes of $\varphi^\P$ 
without introducing any approximations. The lone exception to this, to be 
discussed in Sec.~\ref{mmax convergence}, is that in practice we can truncate 
the number of $m'$ modes at some $|m'|=\mmax$. This is possible because 
the modes fall rapidly with $|m'|$, allowing us to neglect large-$|m'|$ modes 
without introducing significant numerical error.

We must address one more issue before detailing the evaluation of 
Eq.~\eqref{lm' modes}. As discussed in Ref.~\cite{Wardell-Warburton:15}, our 
puncture $\varphi^\P$ is \emph{not} smooth at all points off the particle. The 
particle sits at the north pole $\alpha=0$ of the sphere at $\Delta r=0$, and 
$\varphi^\P$ correctly diverges as $1/\lambda$ there. But even away from the 
particle, for each fixed $\Delta r\neq 0$, $\varphi^\P$ has a directional 
discontinuity at the south pole $\alpha=\pi$, inherited from a directional 
discontinuity in the quantity $\rho$. This  discontinuity is nonphysical. 
$\varphi^\P$ is originally defined from a local expansion in the neighbourhood 
of the particle, but in order to evaluate the integrals~\eqref{lm' modes}, it 
must be extended over the entire sphere spanned by $\alpha^A$. The particular 
discontinuity we face is a consequence of the particular manner in which we 
have performed that extension. Because the total field $\varphi^P+\varphi^\res$ 
is smooth at all points off the particle, this singularity at $\alpha=\pi$ must 
be cancelled by  one in $\varphi^\res$. And because nonsmoothness of a field 
leads to slow falloff with $l$, this discontinuity limits the convergence rate 
of $S_{lm}[\varphi^\res,\varphi^\res]$ and $S_{lm}[\varphi^\res,\varphi^\res]$ 
with $\lmax$. Concretely, the discontinuity introduces terms of the form 
$\frac{(-1)^l}{l}$ into $\varphi^\res_{lm'}$ for all $m'\neq0$.

To eliminate the discontinuity, we must adopt a different extension of 
$\varphi^\P$ over the sphere. Following Ref.~\cite{Wardell-Warburton:15}, we do 
so by introducing a regularizing factor: 
\beq\label{phiP-regularized}
\varphi^\P(\Delta r,\alpha^{A'}) \to {\cal W}_{\sf m}^n(\cos\alpha) 
\varphi^\P(\Delta r,\alpha^{A'}).
\eeq
Here the parameters $n$ and ${\sf m}$ are chosen such that $n\geq k$ and ${\sf m}\geq \mmax$, where $k$ is the order of the puncture and $\mmax$ is the maximum value of $|m'|$ we use. ${\cal W}_{\sf m}^n$'s dependence on these two parameters is dictated by the required behavior at the two poles. To control the behavior at the south pole, we choose a regularizing factor that scales as ${\cal W}_{\sf m}^n = \O[(\pi-\alpha)^{\sf m}]$, which makes ${\cal W}_{\sf m}^n \varphi^\P$  a $C^{\sf m-1}$ function at $\alpha=\pi$. For an otherwise smooth function, standard estimation methods~\cite{Orszag:74} show that this degree of smoothness ensures that the modes $|\varphi^\P_{lm'}|$, and hence $|\varphi^\res_{lm'}|$, fall off as $\lesssim l^{-\sf m\pm 1}$; for sufficiently large ${\sf m}$, this nonspectral decay will be negligible compared to the slow convergence coming from the singularity at the particle. Now, at the same time as satisfying these conditions at the south pole, we must keep control of the behavior at the north pole. Specifically, ${\cal W}_{\sf m}^n$ must leave all $k$ orders intact in the $k$th-order puncture, implying that it must behave as ${\cal W}_{\sf m}^n = 1 + \O(\alpha^n)$ near $\alpha=0$. We satisfy the requirements at both poles by choosing
\begin{align}
{\cal W}_{\sf m}^n &:= 1-\frac{n}{2}\begin{pmatrix}({\sf m} + n - 
2)/2\\ n/2\end{pmatrix} \nonumber\\										
&\quad \times B\!\left(\!\frac{1-\cos\alpha}{2};\frac{n}{2}, \frac{\sf m}{2}\!\right)\!, \label{reg_factor}
\end{align}
where $\begin{pmatrix}p\\q\end{pmatrix}$ is the Binomial coefficient, and $B(z;a,b)$ is the
incomplete Beta function. 

This choice has the required properties at the poles provided $n$ and ${\sf m}$ are positive
integers, and additionally that ${\sf m}$ is even. This is not a significant restriction; as
discussed below, the $\beta$ integrals ensure that only even $m'$ need be considered in our
circular-orbit toy model, and even if this were not the case we could always choose ${\sf m}$
to be the smallest even number greater than $\mmax$. With these restrictions
on $n$ and ${\sf m}$, ${\cal W}_{\sf m}^n$ takes the straightforward form of a
polynomial in $y:=\tfrac{1-\cos\alpha}{2}$, whose coefficients and degree both depend on the
particular choice of $n$ and ${\sf m}$. For example, in all our
computations we use $n=4$ (equal to the highest order of puncture we use) and ${\sf m}=10$ (equal to the value of $\mmax$ we almost exclusively use), in which case ${\cal W}_{10}^4 = 1 - 15 y^2 + 40
y^3 - 45 y^4 + 24 y^5 - 5 y^6$.

Heeding the warnings above about our need for exactness, we must apply this regularization consistently to 
the 4D puncture in all our calculations, not solely in evaluating the 
integrals~\eqref{lm' modes}. So henceforth, we will always use 
Eq.~\eqref{phiP-regularized} as our puncture, with fixed $n$ and ${\sf m}$ independent of the
particular $l,m'$ mode being considered.

With our preparations out of the way, we now describe our evaluation of the 
integrals~\eqref{lm' modes}. We use two methods for computing the double integral~\eqref{lm' modes}, namely
(i) evaluate the $\alpha$ integrals analytically and subsequently evaluate the $\beta$ integrals as numerical elliptic-type integrals, and (ii) 
evaluate both the $\alpha$ and $\beta$ integrals entirely numerically.
The second method is computationally more expensive than the first. However,
we used both methods as an internal consistency check.
We will describe method (i) first and begin by explaining the steps 
in the the analytical evaluation of the $\alpha$ integrals.

\subsubsection{\texorpdfstring{Integration over $alpha$}{Integration over $alphs$}}\label{Barry integration}

We first recall that all of
the $\alpha$ dependence of the puncture~\eqref{eq:singular-field} 
is contained inside the quantity $\rho$. Hence, the integral that we need to evaluate 
takes the general form 
\begin{equation}\label{alphaIntegral}
\int_{-1}^1 \!  {\cal W}_{\sf m}^k(x) P_l^{m'}(x)\rho^{n}\, dx,
\end{equation}
where $x=\cos\alpha$, $P_l^{m'}(x)$ are the associated Legendre polynomials, and $n$ is an odd
integer.

Furthermore, the simple form of ${\cal W}_{\sf m}^n$ as a power series in $\tfrac{1-\cos
\alpha}{2}$ means that we can use Eq.~\eqref{eq:rho} to rewrite it as an even power series in
$\Delta r$ and $\rho$. The integrals \eqref{alphaIntegral} can therefore all be written in the
form
\begin{equation}\label{alphaIntegral-simp}
\int_{-1}^1 \!  P_l^{m'}(x)\rho^{n}\, d x
\end{equation}
for $n$ an odd integer.

Concentrating first on the simplest case of $m'=0$, the integration can be done analytically
using
\begin{widetext}
    \begin{align} \label{eq:alpha-integral}
\int_{-1}^1 (\delta^2 + &1 -x)^{n/2} P^0_\ell(x) \, dx
\nonumber \\
  &= \frac{
    (-1)^{\frac{n+1}{2}}(\delta ^2+2)^{\frac{n}{2}+1}
    \big[\big(\frac{1}{2}\big)_{\frac{n+1}{2}}\big]^2}{\big(l-\frac{n}{2}\big)_{n+2}}
    {}_2F_1(-l,l+1;-\tfrac{n}{2};-\tfrac{\delta^2}{2})
  -\frac{2 \left| \delta \right| \delta^{n+1}}{n+2}
    {}_2F_1 (-l,l+1;\tfrac{n}{2}+2;-\tfrac{\delta ^2}{2})
\nonumber \\
    &=  \frac{
    (-1)^{\frac{n+1}{2}}(\delta ^2+2)^{\frac{n}{2}+1}
    \big[\big(\frac{1}{2}\big)_{\frac{n+1}{2}}\big]^2}{\big(l-\frac{n}{2}\big)_{n+2}}
    \sum_{k=0}^l \frac{(-1)^{k}\delta^{2 k} (l-k+1)_{2 k}}{2^k k! \big(\frac{n}{2}-k+1\big)_k}
  -\left| \delta \right| \delta^{n+1}
    \sum _{k=0}^l \frac{\delta^{2 k} (l-k+1)_{2 k}}{2^k k! \big(\frac{n}{2}+1\big)_{k+1}}.
\end{align}
\end{widetext}
For any given odd integer $n$, this is merely a pair of even polynomials of degree $2 l$ in
$\delta$, one multiplying $(\delta^2+2)^{\frac{p}{2}+1}$ and the other multiplying $|\delta|
\delta^{p+1}$.

Turning to the $m' \ne 0$ case, these can now be written in terms of the $m'=0$ result. Using the
definition for the associated Legendre polynomials in terms of the Legendre polynomials,
\begin{equation}
  P^m_l(x) = (-1)^m (1-x^2)^{m/2} \dfrac{d^m}{dx^m} P_l(x),
\end{equation}
the integral \eqref{alphaIntegral-simp} can be integrated by parts $m'$ times, resulting in an
integral of the form \eqref{eq:alpha-integral} along with a set of $m'$ boundary terms. These
boundary terms are given by
\begin{equation}
  \sum_{k=0}^{m'-1} \Big[ (-1)^k \dfrac{d^k \rho^n}{dx^k} \dfrac{d^{m'-k-1}}{dx^{m'-k-1}} P_l(x) \Big]_{x=-1}^{x=1},
\end{equation}
and are therefore power series in $\delta$ of the same kind as in Eq.~\eqref{eq:alpha-integral}.
The  integrals over $\beta$ then have the same form as for the $m'=0$ case.

\subsubsection{\texorpdfstring{Alternative method for evaluating $\alpha$ integrals}{Alternative method for evaluating alpha integrals}}\label{Jeremy integration}

An alternative, but equivalent strategy for evaluating the $\alpha$ integrals, Eq.~\eqref{alphaIntegral}, is based on expressing ${\cal W}_{\sf m}^n(x)$ and $P_l^{m'}(x)$ as  
finite polynomials in $(1+x)$ and $(1-x)$. For example $n=4$ and ${\sf m}=10$, Eq.~\eqref{reg_factor} can be written as  
\begin{align}\label{WSeries}
{\cal W}_{10}^4(x)
=\frac{3}{16} (1+x)^5-\frac{5}{64} (1+x)^6.
\end{align}
Similarly, for $m\geq0$,
\begin{align}\label{PlmSeries}
P_l^{m}(x)=&\sum^l_{p=0}\sum^{m}_{q=0}c_{lmpq}(1+x)^{p+q-m/2}\nonumber\\ 
&\times(1-x)^{l-p-q+m/2}\,,
\end{align}
where $c_{lmpq}$ are $x$-independent constants given by
\begin{align}
c_{lmpq}&=\frac{(-1)^{m+l-p+q}}{2^l}\binom{l}{\,p\,}^{\!2}\binom{m}{q}\nonumber\\
&\times\frac{(l-p)!}{(l-p-q)!}\frac{p!}{(p-m+q)!}.
\end{align}
Equation~\eqref{PlmSeries} can be derived by using the standard representation $P_l(x)=\frac{1}{2^l}\sum_{p=0}^l\binom{l}{\,p\,}^2(x-1)^{l-p}(x+1)^p$ in the formula $P^m_l = (-1)^m (1-x^2)^{m/2}\frac{d^m}{dx^m}P_l(x)$ and appealing to the Leibniz rule. The analogue of Eq.~\eqref{PlmSeries} for $m<0$ follows from $P_l^{-m}=(-1)^m\frac{(l-m)!}{(l+m)!}P^m_l$; but in practice we need not evaluate the integrals~\eqref{alphaIntegral} for $m'<0$, since for real-valued $\varphi^\P$ we have $\varphi^\P_{l,-m'}=(-1)^{m'}\varphi^{\P*}_{lm'}$

Substituting the polynomials~\eqref{WSeries} and \eqref{PlmSeries} into \eqref{alphaIntegral} yields a sum of integrals of the form $F_{abn}(\delta):=\int^1_{-1}\!\!dx\,(1+x)^{a/2}(1-x)^{b/2}(\delta^2+1-x)^{n/2}$, where $a,b,n$ are positive integers. We write the $\alpha$ integral in Eq.~\eqref{lm' modes} as a linear combination of these integrals $F_{abn}$. Using Wolfram Mathematica, we tabulate analytical formulae for all $F_{abn}$ that appear in this linear combination for $\varphi^\P_{lm'}$ to $l=200$ and $m'=10$. Each of the tabulated formulae is a finite polynomial in $\delta$, and once tabulated, these formulae allow us to almost instantaneously evaluate the $\alpha$ integral. 

\subsubsection{\texorpdfstring{Integration over $\beta$}{Integration over beta}}
We next turn to computing the $\beta$ integrals. The explicit $\beta$-dependent terms in the
puncture, Eq.~\eqref{eq:singular-field}, appear in the form of positive, even powers of
$\sin\beta$. The other dependences on $\beta$ in the integrand appear through $\rho$ (where they
appear as powers of $\chi = 1 - r_0^2 \Omega^2 \sin^2 \beta$), through $\chi$ itself, and through
the factor of $e^{-i m' \beta}$ from the spherical harmonic. With this in mind it can readily be
shown that odd-$m'$ modes vanish and all of the non-vanishing modes are purely real.

Furthermore, following from this structure the net dependence on $\beta$ has two possible forms.
The first term in Eq.~\eqref{eq:alpha-integral} above yields integrals of the form
\begin{equation}
  \int_0^{2\pi} \bigg(2 + \frac{\chi_0 \Delta r^2}{2 r^2_0 \chi}\bigg)^{\frac{n}{2}+1} \chi^{k/2} d\beta,
\end{equation}
where $n$ is an odd integer. For $n=-1$ and $k=-1$ this can be recognized as a complete elliptic
integral of the third kind, with arguments that depend on $\Delta r$, $r_0$, and $\Omega$ (through
$\chi_0$). All other values of $n$ and $k$ can be reduced to this case by integrating by parts a
sufficient number of times. The second type of integral arises from the second term in
Eq.~\eqref{eq:alpha-integral}. This yields integrands involving $\chi^n$ with $n$ an integer; their
integral is a polynomial involving $r_0 \Omega$. Combining these results, we can therefore compute
the integrals over $\beta$ exactly and analytically (in terms of elliptic integrals).

In practice we found it sufficiently efficient (and simpler) to evaluate the $\beta$ integral
directly using numerical integration, rather than manipulting it into elliptic integral form. In
that case, we used the fact that the integrand is symmetric in the sense that
\beq\int^{2\pi}_0 f(\beta)_{lm'} d\beta = 2\int^\pi_0 f(\beta)_{lm'}d\beta\eeq
to reduce the computational cost. To compute the integrals we used a C++ code employing a 15-point
Gauss-Kronrod rule.

\subsubsection{Two-dimension numerical integration}\label{numerical integration}
As a check on our methods, we also evaluated Eq.~\eqref{lm' modes} by computing the double integral
entirely numerically. We used a C++ code employing a 25-point Clenshaw-Curtis integration rule. As
the azimuthal mode number $m'$ increases, the $\beta$ integrals become highly oscillatory,
resulting in loss of accuracy. We found that to improve the accuracy of our results, it was
necessary to split the $\beta$ integral, over the range $[0,\pi]$, into a sum of
$m'$ separate integrals, each over the range $\beta\in[(i-1)/(m'\pi),i/(m'\pi)]$, where $i$ runs
from $1$ to $m'$. In all cases, this fully numerical method agreed with the mixed
analytical-numerical method described above.

\subsection{Calculation of $S_{lm}[\varphi^{\res},\varphi^{\res}]$ and 
$S_{lm}[\varphi^{\res},\varphi^{\P}]$}\label{RR and RS results}

\begin{figure}[tb]
\begin{center}
\includegraphics[width=\columnwidth]{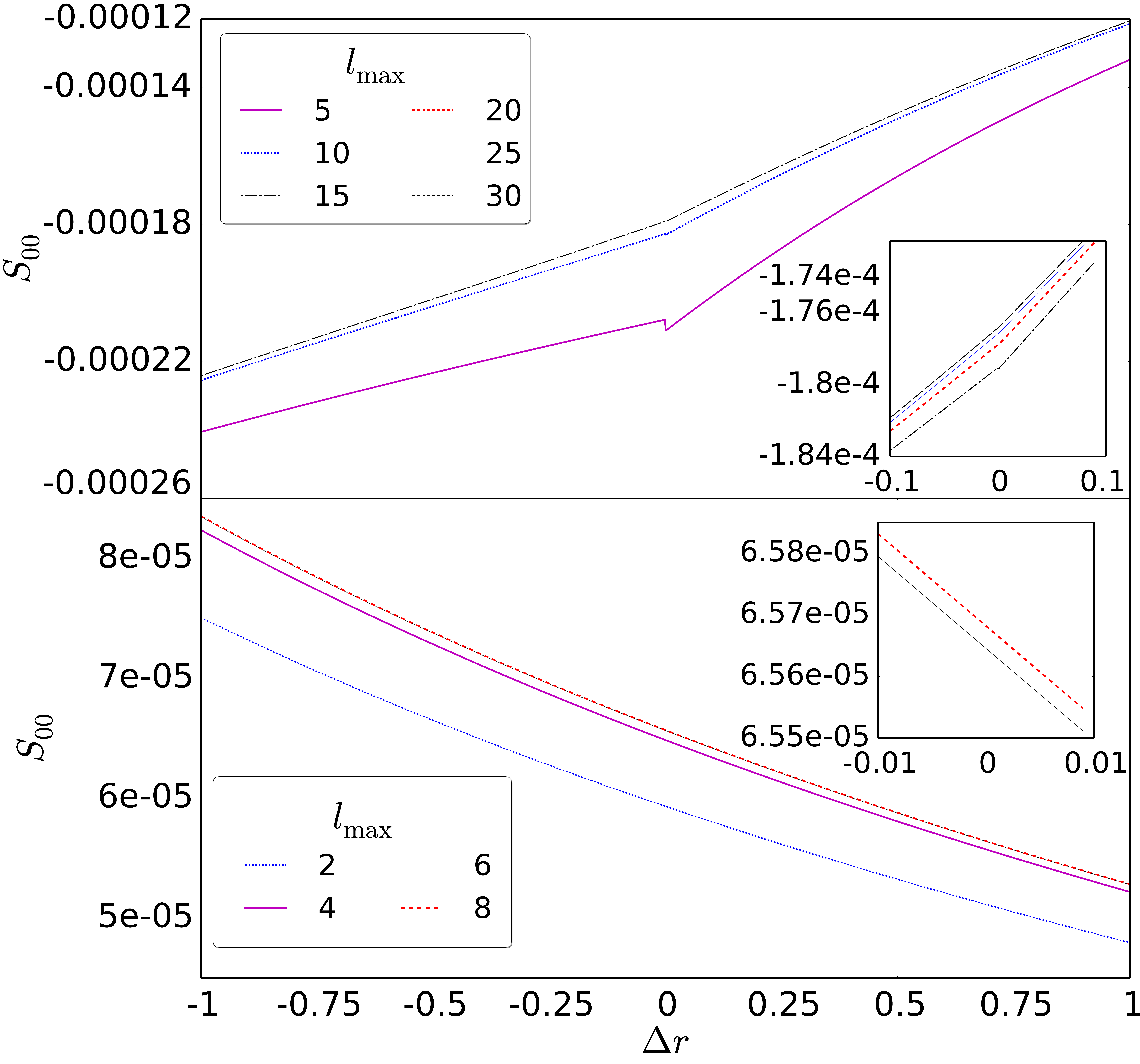}
\caption{\label{Fig:SPR and SRR} Demonstration of rapid convergence of the 
sum~\eqref{Slm} for $S_{00}[\varphi^{\res},\varphi^{\P}]$ (top panel) and 
$S_{00}[\varphi^{\res},\varphi^{\res}]$ (bottom panel). The mode $S_{00}$ is 
plotted as a function of $\Delta r$ for a range of values of $\lmax$. Here we 
use $r_0=10$, $\mmax=10$, and all four orders in the 
puncture~\eqref{phiP}.}
\end{center}
\end{figure}

After obtaining the modes of $\varphi^\P$, we implement the final three steps 
in the strategy outlined at the end of Sec.~\ref{SR strategy}. The results are 
shown in Fig.~\ref{Fig:SPR and SRR} for the monopole modes 
$S_{00}[\varphi^{\res},\varphi^{\P}]$ and 
$S_{00}[\varphi^{\res},\varphi^{\res}]$. We see that unlike 
$S_{lm}[\varphi^{\rm ret},\varphi^{\rm ret}]$, 
$S_{lm}[\varphi^{\res},\varphi^{\P}]$ and 
$S_{lm}[\varphi^{\res},\varphi^{\res}]$ both converge rapidly with increasing 
$\lmax$. On the scale of the main plot, $S_{lm}[\varphi^{\res},\varphi^{\P}]$ has numerically converged by $\lmax=10$ and $S_{lm}[\varphi^{\res},\varphi^{\res}]$ by $\lmax=6$; the insets show the small changes at larger $\lmax$.

However, to make useful predictions about how our strategy extends to 
gravitational fields, we must say more than that it works; we must say 
something about how and when it works. We do this by considering two important 
convergence properties of  Eq.~\eqref{Slm}:
\begin{enumerate}
	\item How quickly do $S_{lm}[\varphi^{\res},\varphi^{\P}]$ and 
$S_{lm}[\varphi^{\res},\varphi^{\res}]$ converge as $\mmax\to\infty$?
	\item How does the convergence of $S_{lm}[\varphi^{\res},\varphi^{\P}]$ 
and $S_{lm}[\varphi^{\res},\varphi^{\res}]$ with $\lmax$ depend on the order of 
the puncture $\varphi^\P$? More pointedly, how high order must the puncture be 
in order to guarantee convergence with $\lmax$? 
\end{enumerate}
The last of these is the most pertinent: as we shall discuss below, if the 
puncture is of too low order, then our strategy simply does not work. However, 
to elucidate that issue, it will be useful to first determine the convergence 
with $\mmax$.

\subsubsection{Convergence with $\mmax$}\label{mmax convergence}
To assess the rate of convergence with $\mmax$, we introduce the finite difference
\beq
\Delta S^{\mmax}_{lm} := S^{\mmax}_{lm}-S^{\mmax-1}_{lm},
\eeq
where $S^{\mmax}_{lm}$ is given by Eq.~\eqref{Slm} with $\varphi^{(1)}_{lm'}$ and $\varphi^{(2)}_{lm'}$ set to zero for $|m'|>\mmax$. Concretely, this means truncating the sum~\eqref{WignerD-rotation} at $|m'|=\mmax$. 

\begin{figure}[t]
\begin{center}
\includegraphics[width=\columnwidth]{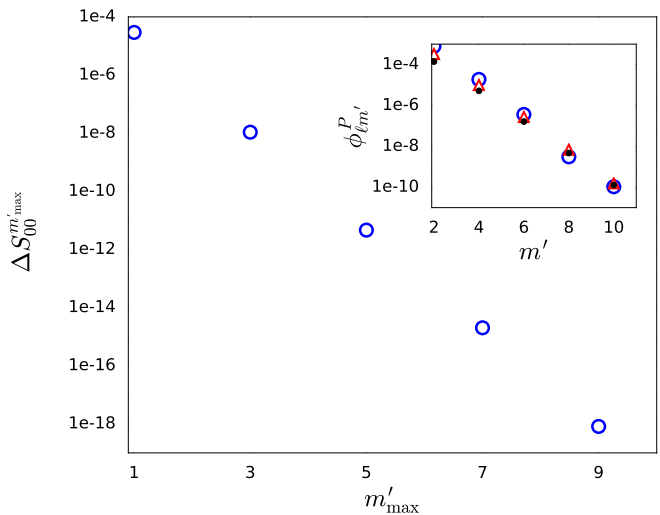}
\caption{\label{Fig:m' convergence} Influence of $m'$ modes on 
$S_{lm}$. The main plot shows $\Delta S^{\mmax}_{00}[\varphi^{\res},\varphi^{\res}]$, which is seen to fall off linearly on the plot's semilog scale, implying exponential decay with $\mmax$. The inset shows $\varphi^\P_{lm'}$ as a function of $m'$ for $l=10$ (open blue circles), $l=20$ (closed black circles), and $l=30$ (open red triangles). In all cases, the modes decay exponentially with $m'$; this behavior carries over to $\varphi^\res_{lm'}$ and explains the falloff of $\Delta S^{\mmax}_{lm}$. To obtain this data we used a fourth-order puncture, $\lmax=30$, and $\Delta r=10^{-4}$.}
\end{center}
\end{figure}

Figure~\ref{Fig:m' convergence} displays the quantity $\Delta S^{\mmax}_{00}[\varphi^{\res},\varphi^{\res}]$ as a function of $\mmax$ at a fixed value of $\lmax$ and $\Delta r$. On the semilogarithmic scale of the plot, $\Delta S^{\mmax}_{00}$ falls linearly, indicating exponential decay.  Although we do not display it, the behavior of $\Delta S^{\mmax}_{00}[\varphi^{\res},\varphi^{\P}]$ is identical, and the behavior is independent of $\Delta r$. Given this rapid decay, we conclude that in practice, we need include only a small number of $m'$ modes; in all other figures in this paper, we use $\mmax=10$.

$S_{lm}$'s rapid convergence with $\mmax$ is a consequence of $\varphi^\P_{lm'}$'s rapid falloff with $m'$. As shown in the inset of Fig.~\ref{Fig:m' convergence}, this falloff is exponential, like that of $\Delta S^{\mmax}_{lm}$. The exponential falloff naturally extends from $\varphi^\P_{lm'}$ to $\varphi^\res_{lm'}$, since $\varphi^{\rm ret}$ will never possess worse convergence properties than $\varphi^\P$, and from there it extends to the convergence of the sum~\eqref{WignerD-rotation} and finally to Eq.~\eqref{Slm}.

We can best understand this behavior, and predict its extension to the gravity case, by obtaining analytical estimates of 
$\varphi^\P_{lm'}$'s falloff. First consider the decomposition into 
$m'$ modes, without the attendant decomposition into $l$ modes. An $m'$ mode is 
defined by $\varphi^\P_{m'}=\int^{2\pi}_0 e^{-im'\beta}\varphi^\P d\beta$. For 
all $\alpha\neq0$, we can integrate by parts $p$ times to express this as
\beq\label{p integrations}
\varphi^\P_{m'} = \left(\frac{-i}{m'}\right)^p\int^{2\pi}_0 
e^{-im'\beta}\partial^p_\beta\varphi^\P d\beta.
\eeq
Hence, 
\beq\label{m' bound}
|\varphi^\P_{m'}| \leq \frac{C(\Delta r,\alpha)}{|m'|^p},
\eeq
where $C(\Delta r,\alpha):=2\pi\max_\beta|\partial^p_\beta\varphi^\P|$ is independent of $m'$. Since $\varphi^\P$ is a $C^\infty$ function of $\beta$ at each fixed $\alpha\neq0,\pi$, 
the bound~\eqref{m' bound} holds for all integers $p\geq0$, and we can see by induction that $\varphi^\P_{m'}$ falls faster than any inverse 
power of $|m'|$. This rate is uniform in $\Delta r$ for each $\alpha\neq0,\pi$; it is not uniform in $(\Delta r, \alpha)$ because the divergence at the particle implies $\sup C(\Delta r,\alpha)=\infty$.

Now consider the  decomposition into $lm'$ modes, which we may write as 
$\varphi^\P_{lm'}=N_{lm'}\int^\pi_0 
\varphi^\P_{m'}P^{m'}_l(\cos\alpha)\sin\alpha d\alpha$, where $N_{lm'}=\sqrt{\frac{2l+1}{4\pi}\frac{(l-m')!}{(l+m')!}}$. Because the the exponential falloff of $\varphi_{m'}$ is nonuniform, we might worry that it does not extend to $\varphi_{lm'}$. However, we can quickly deduce that that is not the case. Using the bound~\cite{Lohofer:98} $|N_{lm'}P^{m'}_l|\leq\sqrt{\frac{2l+1}{8\pi}}$ and Eq.~\eqref{p integrations}, we have
\beq\label{lm' bound}
|\varphi^\P_{lm'}|\leq \frac{1}{|m'|^p}\sqrt{\frac{2l+1}{8\pi}}\int^\pi_0\int^{2\pi}_0|\partial^p_\beta\varphi^\P \sin\alpha|d\alpha.
\eeq
Next we note that $\partial^p_\beta\varphi^\P$ has the same behavior as $\varphi^\P$: it is finite except at $\Delta r=0$, where it diverges as $\sim 1/\alpha$ at small $\alpha$; the derivatives with respect to $\beta$ do not alter this behavior. Hence, the $lm'$-independent integral $\int^\pi_0\int^{2\pi}_0|\partial^p_\beta\varphi^\P \sin\alpha|d\alpha$ exists for all integers $p\geq0$, and we infer by induction that $\varphi^\P_{lm'}$ falls off faster than any power of $|m'|$. Of course, we can only consider large $m'$ if $l$ is at least as large. But because the only $l$ dependence in the bound~\eqref{lm' bound} is the factor $\sqrt{2l+1}$, this consideration does not affect our conclusion.

Of course, exponential convergence does not necessarily mean usefully fast convergence. As we have seen, the falloff of $\varphi^\P_{lm}$ with $l$ is exponentially fast at all points away from $\Delta r=0$, but for practical purposes it is slow for small $\Delta r$. However, that is an artefact of the convergence rate being nonuniform. Crucially, the convergence with $\mmax$ {\em is} uniform in $\Delta r$.

The (uniformly) rapid falloff of $\Delta S^{\mmax}_{lm}[\varphi^\res,\varphi^\P]$ 
and $\Delta S^{\mmax}_{lm}[\varphi^\res,\varphi^\res]$ with $m'_{\rm 
max}$ now follows directly from the rapid falloff of $\varphi^P_{lm'}$.  Because this conclusion relies only on generic behavior of the puncture, it will also apply in the gravity case.

\subsubsection{Convergence with $\lmax$}\label{lmax convergence}
We now turn to the central issue of the convergence rate with $\lmax$. To 
assess that, we examine the finite difference
\beq
\Delta S^\lmax_{lm} := S^{\lmax}_{lm}-S^{\lmax-1}_{lm},
\eeq
where $S^\lmax_{lm}$ is the partial sum in Eq.~\eqref{Slmax}.

\begin{figure}[tb]
\begin{center}
\includegraphics[width=\columnwidth]{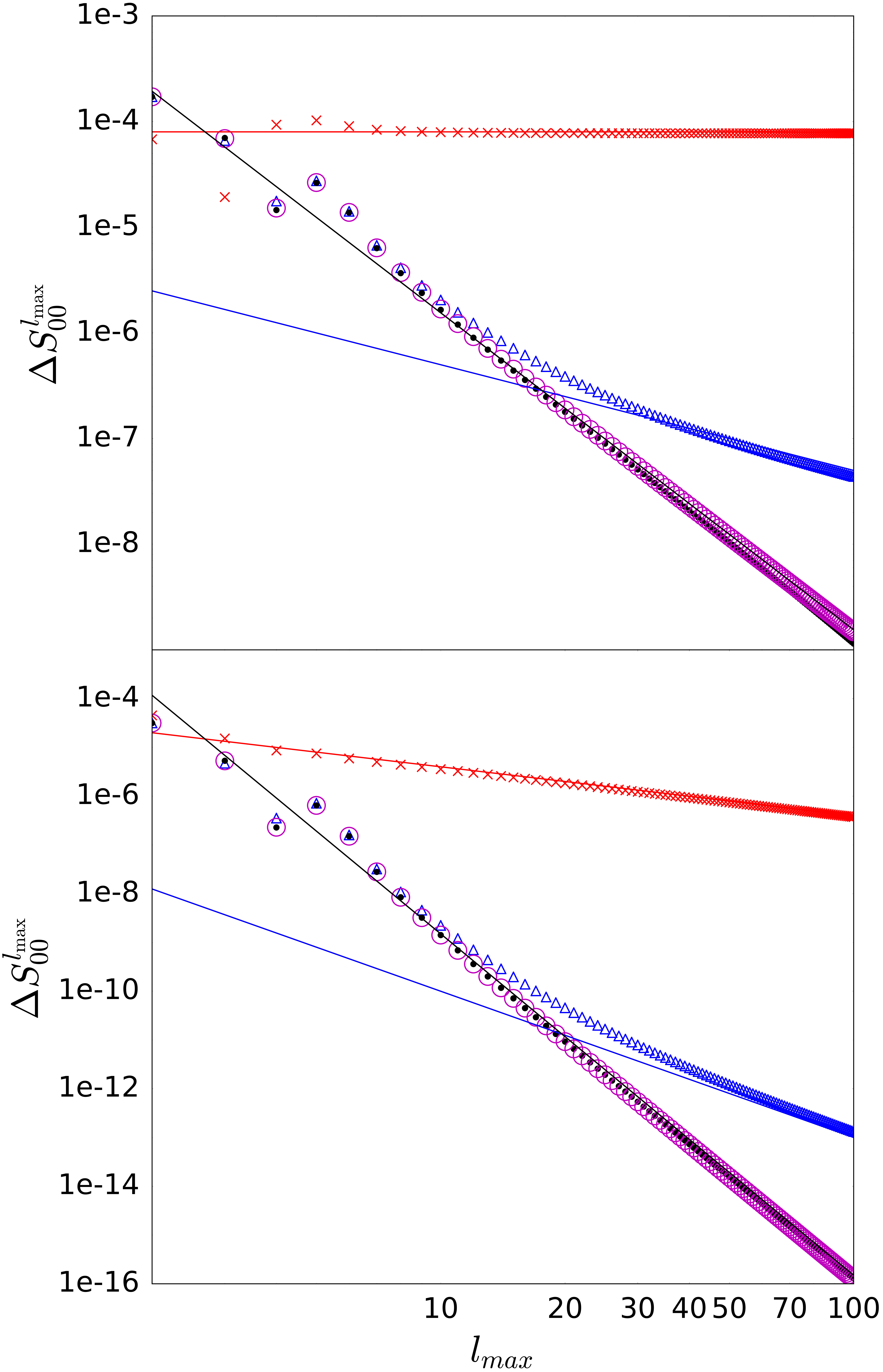}
\caption{\label{Fig: convergence rates} The impact of the puncture order $k$ on $S_{lm}$'s convergence with $\lmax$. $\Delta S^\lmax_{00}[\varphi^{\res},\varphi^{\P}]$ (top panel) and $\Delta S^\lmax_{00}[\varphi^{\res},\varphi^{\res}]$ (bottom) are plotted as functions of $\lmax$. In both panels, results are shown for $k=1$ (red crosses), $k=2$ (blue triangles), $k=3$ (solid black circles), and $k=4$ (open purple circles) and $\Delta r=10^{-12}$. The straight lines show the asymptotic behavior $\propto l^p_{\rm max}$ of the data. In the top panel, listed from top to bottom, they are proportional to $l^0_{\rm max}$, $l^{-1}_{\rm max}$, and $l^{-3}_{\rm max}$; in the bottom panel, $l^{-1}_{\rm max}$, $l^{-3}_{\rm max}$, and $l^{-7}_{\rm max}$. }
\end{center}
\end{figure}

Figure \ref{Fig: convergence rates} displays $\Delta 
S^\lmax_{00}[\varphi^{\res},\varphi^{\P}]$ and $\Delta 
S^\lmax_{00}[\varphi^{\res},\varphi^{\res}]$ at a point very near the particle 
($\Delta r=10^{-12}$). We see that when so close to the particle, the 
sum~\eqref{Slm} exhibits power law convergence. At large enough $\lmax$, this 
will morph into exponential convergence, as $\varphi_{lm}$'s slow exponential decay with $l$  eventually takes over. The further we move  from the 
particle, the less clean the power laws, and the more quickly the exponential 
convergence dominates. 

The most important aspect of the power laws are their dependence on the order 
of the puncture. As we will discuss below, a subtle competition between power laws makes determining the true asymptotics nontrivial, and the numerical results can be misleading. Nevertheless, the numerics provide a useful frame for the discussion. For a $k$th-order puncture, Fig.~\ref{Fig: convergence rates} 
suggests that $S_{00}[\varphi^{\res},\varphi^{\res}]$ converges as
\beq
\Delta S^\lmax_{00}[\varphi^{\res},\varphi^{\res}] \sim 
\begin{cases}l^{-1}_{\rm max} & \text{if } k=1,\\
					l^{-3}_{\rm max} & \text{if } k=2,\\
					l^{-7}_{\rm max} & \text{if $k=3$ or 4};\end{cases}\label{SRR rates}
\eeq
we will demonstrate below that for $k=3$, this inferred falloff is incorrect, and that one would have to go to much larger values of $\lmax$ to see the true asymptotic behavior. But the essential facts are unaltered by that: In order for  $S_{lm}$ to converge with $\lmax$, $\Delta 
S^\lmax_{lm}$ must fall off at least as $l^{-1-p}_{\rm max}$ with $p>0$. Hence, 
to ensure numerical convergence of $S_{00}[\varphi^{\res},\varphi^{\res}]$, we 
must use at least a second-order puncture. Although exponential convergence 
would eventually manifest, in a concrete situation where we have access to modes 
up to $l=\lmax$, the exponential convergence would only assist us at distances 
$|\Delta r|\sim r_0$ from the particle.

Because $\varphi^{\P}$ is singular, $S_{00}[\varphi^{\res},\varphi^{\P}]$ 
converges more slowly than $S_{00}[\varphi^{\res},\varphi^{\res}]$.  According to Fig.~\ref{Fig: convergence rates},
\beq
\Delta S^\lmax_{00}[\varphi^{\res},\varphi^{\P}] \sim \begin{cases}l^{0}_{\rm max} & \text{if } k=1,\\
																					l^{-1}_{\rm max} & \text{if } k=2,\\
																					l^{-3}_{\rm max} & \text{if $k=3$ or 4};\end{cases}\label{SRP rates}
\eeq
again, the inferred falloff for $k=3$ is incorrect. But again, we can nevertheless draw the essential conclusions: Because they are slower than those of Eq.~\eqref{SRR rates}, the falloff rates in Eq.~\eqref{SRP rates} are 
the ultimate determiner of how high order our puncture must be. To ensure 
numerical convergence of 
$S_{00}[\varphi^{\res},\varphi^{\res}]+2S_{00}[\varphi^{\res},\varphi^{\P}]$, 
and hence to allow our overarching strategy to succeed, we must use at least a 
third-order puncture. 

All of the behavior we have just described is generic; it is not particular to 
the monopole. We  now  argue, by way of scaling estimates for arbitrary $k$, that it also extends to the gravitational case. As a byproduct of our derivation, we will also discover, as alluded to above, that the power laws in Eqs.~\eqref{SRR rates} and \eqref{SRP rates} are not the true asymptotic falloffs for $k=3$.

First let us continue to focus on $S_{00}$. We will afterward generalize  to 
arbitrary $lm$. Although in practice we use Eq.~\eqref{Slm} to compute 
$S_{lm}$, Eq.~\eqref{Slm'} will be more useful for our argument. For $l=0$, 
Eq.~\eqref{coupling} simplifies to
\beq
C^{000}_{l_1m_1s_1l_2m_2s_2} = 
\frac{(-1)^{m_1+s_1}}{\sqrt{4\pi}}\delta^{l_1}_{l_2}\delta^{m_1}_{-m_2}\delta^{s_1}_{-s_2},
\eeq
where $\delta^i_j$ is a Kronecker delta. Substituting this into 
Eq.~\eqref{Slm'} and simplifying, we find
\begin{align}\label{S00}
S_{00} &=  \frac{1}{\sqrt{4\pi}}\sum_{lm'}\!\bigg[\partial_r \varphi^{(1)}_{lm'}\partial_r \varphi^{(2)*}_{lm'} +\frac{l(l+1)}{r^2}\varphi^{(1)}_{lm'}\varphi^{(2)*}_{lm'}\nonumber\\
		&\quad +\frac{\Omega^2}{4}(\mu^-_{lm'} \varphi^{(1)}_{l, m' + 1} - \mu^+_{lm'} \varphi^{(1)}_{l, m' - 1}) \nonumber\\
		&\quad \times(\mu^-_{lm'} \varphi^{(2)*}_{l, m' + 1} -  \mu^+_{lm'} \varphi^{(2)*}_{l, m' - 1})\bigg].
\end{align}
Based on the result that $\varphi_{lm'}$ 
decays exponentially with $m'$, we may disregard the sum over $m'$ for the 
purpose of finding the scaling with $\lmax$. We then obtain the estimate
\begin{align}\label{DS00}
\Delta S^\lmax_{00} &\sim  
			\partial_r \varphi^{(1)}_{\lmax 0'}\partial_r 
\varphi^{(2)}_{\lmax 0'} +l_{\rm 
max}^2\varphi^{(1)}_{\lmax0'}\varphi^{(2)}_{\lmax0'}.
\end{align}
Note that the $t$ derivatives in the original source simply contribute to the second term here. They appear in Eq.~\eqref{S00}  as the term proportional to $\Omega^2$, the dominant piece of which  is given by $\tfrac{1}{2}\Omega^2 l(l+1) \varphi^{(1)}_{l0'}\varphi^{(2)}_{l0'}$.

We now appeal to standard results for the large-$l$ behavior of 
$\varphi^\P_{l0'}$ and $\varphi^\res_{l0'}$~\cite{Heffernan-etal:12}. It is well known that when evaluated on the particle, (a) $\partial^n_r\varphi^\P_{l0'}Y_{l0'}\sim l^n$ and $\partial^n_r\varphi^\res_{l0'}Y_{l0'}\sim l^{n-k}$ for a $k$th-order puncture, and (b) the odd negative powers of $l$ in $\partial^n_r\varphi^\res_{l0'}Y_{l0'}$ identically vanish. Noting that $Y_{l0'}(0,\beta)\sim l^{1/2}$, we infer that $\varphi^\P_{l0'}\sim l^{-1/2}$, 
$\partial_r\varphi^\P_{l0'}\sim l^{1/2}$, $\varphi^\res_{l0'}\sim 
l^{-5/2-2\lfloor \frac{k-1}{2}\rfloor}$, and $\partial_r\varphi^\res_{l0'}\sim 
l^{-1/2-2\lfloor \frac{k}{2}\rfloor}$, where $\lfloor s\rfloor$ denotes the 
largest integer less than or equal to $s$. These results hold at $\Delta r=0$; 
at finite $\Delta r$, they transition into exponential decay in the now 
familiar manner. Substituting this behavior into Eq.~\eqref{DS00} yields
\begin{subequations}
\begin{align}
\Delta S^\lmax_{00}[\varphi^\res,\varphi^\res] &\sim  
			l_{\rm max}^{-1-4\lfloor \frac{k}{2}\rfloor} +l_{\rm max}^{-3-4\lfloor \frac{k-1}{2}\rfloor\label{DS00RR-competition}}\\
			&\sim l_{\rm max}^{1-2k}\label{DS00RR}
\end{align}
\end{subequations}
and
\begin{subequations}
\begin{align}
\Delta S^\lmax_{00}[\varphi^\res,\varphi^\P] &\sim  
			l_{\rm max}^{-2\lfloor \frac{k}{2}\rfloor} +l_{\rm max}^{-1-2\lfloor \frac{k-1}{2}\rfloor}\label{DS00RP-competition}\\
			&\sim l_{\rm max}^{1-k}.\label{DS00RP}
\end{align}
\end{subequations}
In Eqs.~\eqref{DS00RR-competition} and \eqref{DS00RP-competition}, the first term arises from $(\partial_r\varphi)^2$ and the second arises from $(\partial_t\varphi)^2+\frac{1}{r^2}\partial_A\varphi\partial^A\varphi$; these two terms alternate in dominance from one $k$ to the next.

To extend our estimates to generic $lm$ modes, we note that in 
Eq.~\eqref{Slmax}, when $l_1\sim\lmax\gg l$, the triangle inequality also 
enforces $l_2\sim\lmax\gg l$. We can then appeal to the approximation 
\beq
\begin{pmatrix}l & l_1 & l_2 \\ m & m_1 & m_2\end{pmatrix} \approx 
(-1)^{l_2+m_2}\frac{d^{l}_{m,l_2-l_1}(\gamma)}{\sqrt{l_1+l_2+1}}\sim \frac{1}{l_{\rm max}^{1/2}}
\eeq
for $l\ll l_1,l_2$, where $\cos\gamma = (m_1-m_2)/(l_1+l_2+1)$. This implies
\beq
C^{lms}_{l_1m's_1l_2m_2s_2}\sim l_{\rm max}^0.
\eeq
Given this, we can apply the same arguments as above and find the same scaling 
estimates: $\Delta S^\lmax_{lm}[\varphi^\res,\varphi^\res]\sim  l_{\rm 
max}^{1-2k}$ and $\Delta S^\lmax_{lm}[\varphi^\res,\varphi^\P]\sim  l_{\rm 
max}^{1-k}$. From this, we again conclude that at least a third-order puncture is needed to ensure convergence.


\begin{figure}[tb]
\begin{center}
\includegraphics[width=\columnwidth]{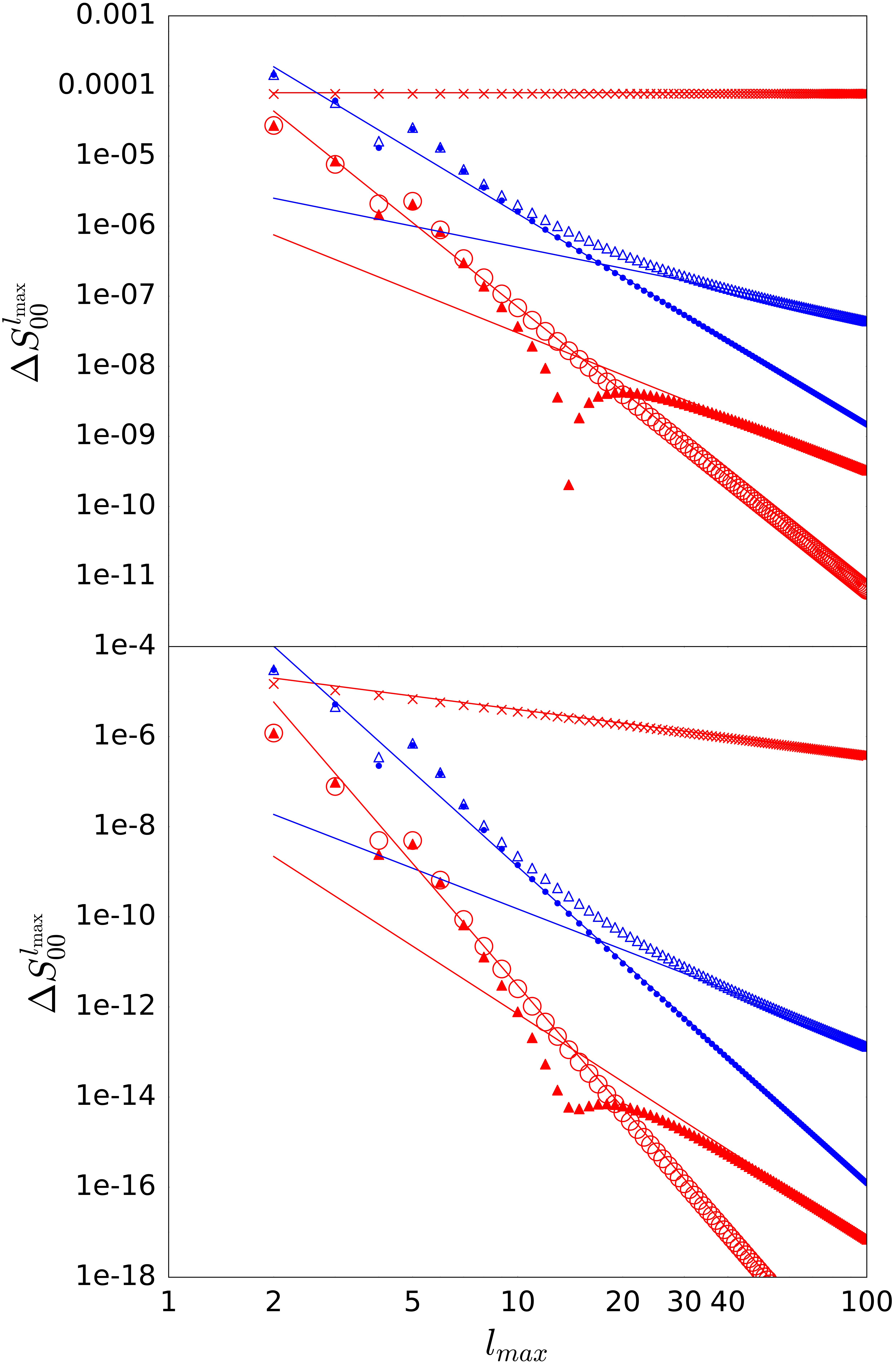}
\caption{\label{Fig: competition} Comparison of two contributions to $\Delta S^\lmax_{00}[\varphi^{\res},\varphi^{\P}]$ (top panel) and $\Delta S^\lmax_{00}[\varphi^{\res},\varphi^{\res}]$ (bottom), using the same parameters as in Fig.~\ref{Fig: convergence rates}. Red symbols represent the contribution from $\partial_r \varphi^{(1)}\partial_r \varphi^{(2)}$ in Eq.~\eqref{unrotated S}, and blue symbols, the contribution from $\partial_t \varphi^{(1)} \partial_t\varphi^{(2)}  + \frac{1}{r^2}\Omega^{AB}\partial_A\varphi^{(1)}\partial_B\varphi^{(2)}$. For the red symbols, crosses correspond to $k=1$, solid triangles to $k=2$ and $k=3$, and open circles to $k=4$; for the blue symbols, open triangles correspond to $k=1$ and $k=2$, and solid circles to $k=3$ and $k=4$. 
 The reference lines are proportional to $l^0_{\rm max}$, $l^{-1}_{\rm max}$, $l^{-2}_{\rm max}$, $l^{-3}_{\rm max}$, and $l^{-4}_{\rm max}$ in the top panel, and to $l^{-1}_{\rm max}$, $l^{-3}_{\rm max}$, $l^{-5}_{\rm max}$, $l^{-7}_{\rm max}$, and $l^{-9}_{\rm max}$ in the bottom panel. For $k=1$, the dominant contribution comes from the red crosses; for $k=2$, the open blue triangles; for $k=4$, the solid blue circles. For $k=3$, the dominant contribution appears to come from the solid blue circles, but because the solid red triangles are falling more slowly, they will eventually become dominant at sufficiently large $\lmax$.}
\end{center}
\end{figure}

We now return to the numerically determined scalings in Eqs.~\eqref{SRR rates} and \eqref{SRP rates}. Comparing them to Eqs.~\eqref{DS00RR} and 
\eqref{DS00RP}, we see that the numerical estimates agree with the analytical ones except in the case of $k=3$, as mentioned previously. This discrepancy stems from Eqs.~\eqref{DS00RR-competition} and \eqref{DS00RP-competition}. There we see that for a given $k$, two power laws compete for dominance. In practice, we find that the coefficients of these power laws can dramatically differ. Let us focus on $\Delta S^\lmax_{00}[\varphi^\res,\varphi^\res]$ for concreteness. For $k=3$, the dominant power in Eq.~\eqref{DS00RR-competition} is $l^{-5}_{\rm max}$, and it arises from $(\partial_r\varphi)^2$; the subdominant power is $l^{-7}_{\rm max}$, and it arises from $(\partial_t\varphi)^2+\frac{1}{r^2}\partial_A\varphi\partial^A\varphi$. In our numerical results, we only see the latter, subdominant behavior. Why? Because it comes with an enormously larger numerical coefficient. This is demonstrated in Fig.~\ref{Fig: competition}, which plots the contributions from $(\partial_r\varphi)^2$ and $(\partial_t\varphi)^2+\frac{1}{r^2}\partial_A\varphi\partial^A\varphi$ separately. Each of the separate terms is in agreement with Eqs.~\eqref{DS00RR-competition} and \eqref{DS00RP-competition}, but we see that for $k=3$, $\Delta [(\partial_r\varphi)^2]^\lmax_{00}$ is hugely suppressed relative to $\Delta[(\partial_t\varphi)^2+\frac{1}{r^2}\partial_A\varphi\partial^A\varphi]^\lmax_{00}$, even though $\Delta[(\partial_r\varphi)^2]^\lmax_{00}$ is decaying more slowly. In fact, by fitting the curves, we can estimate that for $k=3$ and $r_0=10$, the true asymptotic behavior would only become numerically apparent at $\lmax>450$.

This competition between terms appears to be a robust feature of the model: numerical investigations show that it is independent of $l$ and $m$ and largely independent of $r_0$, though it subsides at smaller values of $r_0$. Furthermore, the underlying cause is not confined to $k=3$, as we find that the coefficients of various powers of $1/\lmax$ in $\Delta S^\lmax_{lm}$ often differ by factors of $10^4$ or more. Indeed, this is true not just in $\Delta S^\lmax_{lm}$, but also within the individual contributions $\Delta [(\partial_r\varphi)^2]^\lmax_{lm}$, $\Delta[(\partial_t\varphi)^2]^\lmax_{lm}$, and $\frac{1}{r^2}[\partial_A\varphi\partial^A\varphi]^\lmax_{lm}$. We have no reason to believe that this is particular to our model. Wildly disparate coefficients of the powers of $1/\lmax$ could very well occur in the gravitational case as well. Because of this, in principle, one might encounter a situation in which one's numerical results had {\em appeared} to converge, when in fact a divergent power of $1/\lmax$ was still waiting to emerge at larger $\lmax$. One can only eliminate this possibility by appealing to analytical estimates of the sort in Eqs.~\eqref{DS00RR} and \eqref{DS00RP}.

With this additional impetus, we now extend our estimates to the gravitational case. Because $\delta^2G_{ilm}$ has the same form as $S_{lm}$, and because 
$h^{1\res}_{ilm'}$ and $h^{1\P}_{ilm'}$ have the same behavior as 
$\varphi^\P_{lm'}$ and $\varphi^\res_{lm'}$, similar estimates will apply. 
The only difference between the two cases is that $\delta^2G$ contains 
terms of the form $h\partial^2 h$ and terms that mix $t,r,\theta^A$ derivatives. Assume we can account for these changes by adopting a generic form
\begin{align}
\Delta \delta^2G^\lmax_{ilm} &\sim  
			\partial_r h_{j\lmax 0'}\partial_r h_{k\lmax 0'} +\lmax^2h_{j\lmax0'}h_{k\lmax0'} \nonumber\\
			&\quad + \lmax h_{j\lmax0'}\partial_rh_{k\lmax0'}\nonumber\\
			&\quad +h_{j\lmax0'}\partial^2_rh_{k\lmax0'}
\end{align}
in place of Eq.~\eqref{DS00}. Using $\partial^2_rh^\P_{il0'}\sim l^{3/2}$, $\partial^2_rh^\res_{il0'}\sim l^{1/2}$ for $k=1$, $\partial^2_rh^\res_{il0'}\sim l^{-1/2-2\lfloor\frac{k-1}{2}\rfloor}$ for $k>1$, and the scalings given above for the lower derivatives, we find that $\Delta \delta^2G^\lmax_{ilm}[h^\res,h^\P]\sim \lmax^{1-k}$ and $\Delta \delta^2G^\lmax_{ilm}[h^\res,h^\res]\sim \lmax^{-k-2\lfloor\frac{k-1}{2}\rfloor}$. The first of these convergence rates is the slower of the two, and it is identical to the scalar model. 
Therefore, we conclude that like in the scalar model, for 
our strategy to be effective in the gravitational case, \emph{it requires at 
least a third-order puncture $h^{1\P}_{\mu\nu}$}.

\section{Computing $S_{lm}[\varphi^{\P},\varphi^{\P}]$}\label{SS}
The only term that remains to be computed in Eq.~\eqref{S pieces} is 
$S_{lm}[\varphi^\P,\varphi^P]$. As we described in the outline of our strategy, 
we calculate the modes of $S_{lm}[\varphi^\P,\varphi^P]$ by substituting the 4D 
expression~\eqref{phiP-regularized} into the 4D expression for $S$ and then 
integrating against spherical harmonics to obtain the modes.

More precisely, our procedure is  summarized by the following four steps:
\begin{enumerate}
	\item Begin with the puncture field~\eqref{phiP-regularized} in the 
rotated coordinates $\alpha^{A'}$. 
	\item Construct the 4D expression $S[\varphi^{\P},\varphi^{\P}]$ in 
$\alpha^{A'}$ coordinates using Eq.~\eqref{rotated S body}.
	\item Decompose $S[\varphi^{\P},\varphi^{\P}]$ into $lm'$ modes 
$S_{lm'}[\varphi^{\P},\varphi^{\P}]$ by evaluating the integrals~\eqref{Slm' 
modes}.
	\item Use Eq.~\eqref{WignerD-rotation} to obtain the $lm$ modes  
$S_{lm}[\varphi^{\P},\varphi^{\P}]$.
\end{enumerate}

The nontrivial step in this procedure is the evaluation of the 
integrals~\eqref{Slm' modes}. We perform that evaluation in the same manner as 
we did the integrals in Sec.~\ref{phiPlm computation}. Again we use two 
independent methods of evaluation: fully numerical and mixed 
analytical-numerical. The only new features of the integrals is that the integrand now contains explicit factors of $\sin\alpha$ and $\cos\alpha$
as well as higher powers, and even powers, of $\rho$ in their denominator. Because Eq.~\eqref{eq:alpha-integral} is defined only for odd $n$, the method described in Sec.~\ref{Barry integration} is not immediately applicable; an even-$n$ analog of Eq.~\eqref{eq:alpha-integral} would be required. However, the even powers of $n$ are readily 
handled by the methods described in Secs.~\ref{Jeremy integration} and \ref{numerical integration}. 


After performing the integrals, we arrive at our promised result displayed in 
Fig.~\ref{Fig:together}. There we see that near the particle, where 
$S_{lm}[\varphi^{\rm ret},\varphi^{\rm ret}]$ converges too slowly with $\lmax$ 
to see any singularity at $\Delta r=0$, our computed $S_{lm}$ correctly behaves 
as $1/(\Delta r)^2$. Further from the particle, where $S_{lm}[\varphi^{\rm 
ret},\varphi^{\rm ret}]$ rapidly converges with $l_{\rm max}$, our computed 
$S_{lm}$ correctly recovers $S_{lm}[\varphi^{\rm ret},\varphi^{\rm ret}]$.

\section{Conclusion}\label{Conclusion}

We have now demonstrated that our strategy successfully circumvents the problem 
of slow convergence described in the introduction. This success is encapsulated 
by Fig.~\ref{Fig:together}. 

The core tools in our strategy are adopted from mode-sum regularization and 
effective-source schemes, but our analysis has highlighted several unforeseen 
complications in applying these standard methods. Specifically, we have found 
that notable intricacies arise in computing mode decompositions in rotated 
coordinates that place the particle at the north pole. Traditionally, the time 
dependence of the rotation could be treated cavalierly, but in the calculations 
described here, it must be handled with care; traditionally, only one azimuthal 
mode (or a specific few~\cite{Warburton-Wardell:14,Wardell-Warburton:15}) are 
required in the rotated coordinates, but here a significant number must be 
computed; and traditionally, the relevant Legendre integrals can often be 
simplified by analyzing them in the limit $r\to r_0$, but here they must be 
evaluated \emph{exactly} in some finite range of $r$ around 
$r_0$.

Although our implementation has been in a simple scalar toy model, our strategy 
and computational tools are not in any way specific to that model, and they can 
be applied directly to the physically relevant gravitational problem. For 
example, for a particle in a Schwarzschild background, the steps involved in 
that calculation are as follows:
\begin{enumerate}
\item Begin with two ingredients: 
	\begin{enumerate} 
	\item numerically computed tensor-harmonic modes $h^1_{ilm}$ of the 
first-order retarded field in the unrotated coordinates $(t,r,\theta^A)$,
	\item a 4D expression for the puncture $h^{1\P}_{\mu\nu}$ in the 
rotated coordinates $(t,r,\alpha^{A'})$. 
	\end{enumerate}
	For a given numerical accuracy target, the higher the order of the 
puncture, the fewer modes $h^1_{ilm}$ are required; correspondingly, the more 
modes of $h^1_{ilm}$ are computed, the lower the necessary order of the 
puncture. However, following the discussion in Sec.~\ref{RR and RS results}, 
the puncture must be of at least third  order (counting the leading, 
one-over-distance term as first order). 
\item Using the coupling formula~\eqref{ddGilm}, given explicitly in 
Ref.~\cite{Pound:16}, compute the modes $\delta^2 G_{ilm}[h^1,h^1]$. They 
should be computed over the entire numerical domain except in a region ${\cal 
R}=[r_0-a,r_0+b]$ around the particle, choosing ${\cal R}$ such that it 
contains all points at which the sums in Eq.~\eqref{ddGilm} fail to numerically 
converge.
\item In the region ${\cal R}$, compute the tensor-harmonic modes 
$h^{1\P}_{ilm'}$ in the rotated system and then use Wigner D matrices to obtain 
the modes $h^{1\P}_{ilm}$ in the unrotated system, as described in 
Sec.~\ref{phiPlm computation}. From the result, compute the modes 
$h^{1\res}_{ilm}=h^1_{ilm}-h^{1\P}_{ilm}$ of the residual field.
\item Using the coupling formula~\eqref{ddGilm}, compute the modes $\delta^2 
G_{ilm}[h^{1\P},h^{1\res}]$ and $\delta^2 G_{ilm}[h^{1\res},h^{1\res}]$ in 
${\cal R}$.
\item Following the treatment of time derivatives in the Appendix, express 
$\delta^2 G_{\mu\nu}[h^{1\P},h^{1\P}]$ in the rotated coordinates 
$(t,r,\alpha^{A'})$. In ${\cal R}$, compute the modes $\delta^2 
G_{ilm}[h^{1\P},h^{1\P}]$ in the same manner that one computed $h^{1\P}_{ilm}$.
\item Sum the results  $\delta^2 G_{ilm}[h^{1\P},h^{1\P}]+2\delta^2 
G_{ilm}[h^{1\P},h^{1\res}] +\delta^2 G_{ilm}[h^{1\res},h^{1\res}]$ to obtain 
the complete $\delta^2 G_{ilm}$ in the region ${\cal R}$. Combined with the 
result from step 2, this provides $\delta^2 G_{ilm}$ everywhere in the 
numerical domain.
\end{enumerate}
This general procedure would also apply to any other nonlinear perturbative 
problem containing localized singularities, so long as (i) one wished to decompose the problem into harmonics (or some set of orthogonal polynomials) and (ii) one had access to a local, non-decomposed approximation to the singularity.

We recently reported~\cite{Pound:16b} how the strategy presented here has been combined with those 
developed in Refs.~\cite{Warburton-Wardell:14,Wardell-Warburton:15,Pound:15c} 
to compute second-order self-force effects on quasicircular orbits in 
Schwarzschild spacetime. A future paper will describe that calculation in detail.

\acknowledgments
We thank Leor Barack and Niels Warburton for helpful discussions. J.M. and A.P. 
acknowledge support from the European Research Council under the European 
Union's Seventh Framework Programme (FP7/2007-2013)/ERC Grant No. 304978.
B.W. was supported by the Irish Research Council, which is
funded under the National Development Plan for Ireland. This material is based
upon work supported by the National Science Foundation under Grant Number 1417132.

\appendix

\section{Rotations}\label{rotations}
In Sec.~\ref{SS}, we require a 4D representation of 
$S=t^{\mu\nu}\partial_\mu\varphi^\P_1\partial_\nu\varphi^\P_1$, given only the 
expression~\eqref{phiP} for $\varphi^\P_1$, an expression written in a 
coordinate system in which the particle is instantaneously at the north pole. 
This is nontrivial because there is no explicit time dependence in 
Eq.~\eqref{phiP},\footnote{This fact is specific to circular orbits. For 
noncircular orbits, even in these rotated coordinates, $\varphi^\P_1$ would 
depend on time through its dependence on the orbital radius $r_p(t)$.} making 
it unclear how to evaluate the  $t$ derivatives in $S$. Here we consider two 
ways of tackling this problem: via a time-dependent rotation and via a 
one-parameter family of rotations. We will refer to the first as the 4D method, 
the second as the 2D method. To assist the discussion, we split the unrotated coordinates into $x^\mu=(x^a,\theta^A)$, where $x^a=(t,r)$ and $\theta^A=(\theta,\phi)$, thereby splitting the manifold into the Cartesian product ${\cal M}^2\times S^2$, where ${\cal M}^2$ is the $x^a$ plane and $S^2$ is the unit sphere.
 
In the first approach, we would use a 4D coordinate transformation $x^\mu\to 
x^{\mu'}=(x^{a'},\alpha^{A'})$ given by $x^{a'}=x^a$ and $\alpha^{A'}=\alpha^{A'}(\theta^A,t)$, where 
$\alpha^{A'}=(\alpha,\beta)$, such that at each 
fixed $t$, the transformation would be a 2D rotation that placed the particle 
at the north pole.
In this case, all tensors would transform in the usual 4D way, including 
tensors tangent ${\cal M}^2$; the transformation mixes ${\cal M}^2$ with $S^2$. For example, for a dual vector $w_\mu$ we 
would have $w_{t}\to w_{t'}=w_t+\dot\theta^A w_A$, $w_r\to w_{r'}=w_r$, and 
$w_A\to w_{A'}=\Omega^A\sub{A'}w_A$, where
\begin{align}
\dot\theta^A &:= \frac{\partial\theta^{A}}{\partial t'},\\
\Omega^A\sub{A'} &:= \frac{\partial \theta^A}{\partial\alpha^{A'}}.
\end{align}
In the coordinates $x^{\mu'}$, the particle would be permanently at the north 
pole, with four-velocity $u^{a'}=u^a$ and $u^{A'}=0$. [Since the coordinates 
are singular at the particle's position at the north pole, $u^{A'}$ is not 
strictly well defined. But if we introduce local Cartesian coordinates 
$x^{i'}=(r_0 \alpha\cos\beta,r_0 \alpha\sin\beta)$, then we can establish 
$u^{i'}=0$, allowing us to freely set $u^{A'}=0$.] In this method, all 
components would be expressed in the primed coordinate system, meaning the only 
time derivatives appearing in $S$ would be $\partial_{t'}\phi^\P_1$. For 
circular orbits, these derivatives would trivially vanish because $\phi^\P_1$ 
contains no explicit dependence on $t'$; the $t$ dependence would be entirely 
encoded in the transformation law's dependence on $\dot\theta^A$. 

Although the 4D method is practicable, we henceforth adopt the second, 2D 
method, for reasons described below. In this approach, instead of a 4D 
coordinate transformation, we consider a different 2D rotation at each instant 
of $t$. We may write this as $\alpha^{A'}_{t}=\alpha^{A'}(\theta^A,t)$. This is 
superficially the same as the 4D method, but the time at which the rotation is 
performed is now a parameter of the rotation rather than a coordinate, and for 
each value of the parameter, we have a different coordinate system; for 
example, if the rotation is performed at time $t_0$, it induces a coordinate 
system $(t,r,\alpha^{A'}_{t_0})$. Because the transformation is restricted to 
$S^2$, tensors tangent to ${\cal M}^2$ transform as scalars and those tangent 
to $S^2$ transform as tensors on $S^2$: for the same dual vector $w_\mu$ 
mentioned above, we now have $w_{a}\to w_a$ and $w_A\to 
w_{A'}=\Omega^A\sub{A'}w_A$. Unlike in the 4D method, where the particle was 
permanently at the north pole, here it is only there at the particular instant 
at which the rotation is performed, with an instantaneous four-velocity 
$(u^a,u^{A'})=(u^a,u^\phi,0)$ at that time. [As above, this value of $u^{A'}$ 
comes from consideration of the locally Cartesian components, which can be 
established to be $u^{i'}=(r_0u^\phi,0)$.] Time derivatives in this method are 
evaluated as derivatives with respect to the parameter $t$: 
$\partial_{t}\phi^\P_1 = \dot\alpha^{A'}\partial_{A'}\phi^\P_1$, where
\beq\label{alphadot}
\dot \alpha^{A'}:=\frac{\partial \alpha^{A'}}{\partial 
t}=-\Omega^{A'}\sub{A}\dot\theta^A.
\eeq
Here 
$\Omega^{A'}\sub{A}:=\frac{\partial\theta^{A'}}{\partial\theta^A}=(\Omega^{A}
\sub{A'})^{-1} = \Omega^{A'B'}\Omega_{AB}\Omega^B\sub{B'}$, and the second 
equality in Eq.~\eqref{alphadot} follows from the implicit function theorem.

In our toy model, the above two methods both lead to the 
result
\beq\label{rotated S}
S=(\partial_r\varphi^P)^2 + 
(r^{-2}\Omega^{A'B'}+\dot\alpha^{A'}\dot\alpha^{B'})\partial_{A'}
\varphi^P\partial_{B'}\varphi^P.
\eeq
However, in gravity the two methods would lead to quite different calculations 
when performing decompositions into tensor harmonics. 
Furthermore, only the 2D method is immediately applicable to the decomposition 
strategy of Ref.~\cite{Wardell-Warburton:15}.\footnote{To see this, consider 
$\delta^2 G_{\mu\nu}[h^{1\P},h^{1\P}]$. In the strategy used in 
Ref.~\cite{Wardell-Warburton:15}, as in our 2D method described here, a 
quantity such as $\delta^2 G_{tt}$ is treated as a scalar, that scalar is then 
written in terms of the coordinates $\alpha^{A'}$, and it is decomposed into 
scalar harmonics by integrating against $Y_{lm}(\alpha^{A'})$. Contrary to 
this, in the 4D method, the scalar-harmonic decomposition of $\delta^2 G_{tt}$ 
would be constructed from the scalar, vector, and tensor-harmonic decompositions 
of $\delta^2G_{t't'}$, $\delta^2 G_{t'A'}$, and $\delta^2 G_{A'B'}$, using the 
transformation $\delta^2G_{tt}=\delta^2 G_{t't'}+2\dot\alpha^{A'}\delta^2 
G_{t'A'}+\dot\alpha^{A'}\dot\alpha^{B'}\delta^2 G_{A'B'}$.} Hence, the 2D 
method is preferred here.

All of the above is fairly general. When we specialize to our particular case 
of circular orbits with frequency $\Omega$, the transformation is given by
\begin{align}
\theta &= \arccos(\sin\alpha\sin\beta),\\
\phi &= \arccos\{\cos\alpha/\sin[\arccos(\sin\alpha\sin\beta)]\}+\Omega t,
\end{align}
which implies  $(u^a,u^{A'})=u^t(1,0,\Omega,0)$ and
\begin{align}
\dot\theta^A &= (0,\Omega),\\
\dot\alpha^{A'} &= \Omega(-\cos\beta,\cot\alpha\sin\beta).\label{alpha dot}
\end{align}
The final expression for $S$, used in our computations in Sec.~\ref{SS}, is 
given by Eq.~\eqref{rotated S} with Eq.~\eqref{alpha dot}.


\bibliography{bibfile}

\end{document}